\documentclass[11pt]{article}

\usepackage{geometry}
\usepackage[utf8]{inputenc}
\usepackage[T1]{fontenc}
\usepackage{amsmath,amssymb,amsfonts}	
\usepackage{graphicx}
\usepackage{color}
\usepackage{xcolor}
\usepackage{xspace}
\usepackage{booktabs}
\usepackage{slashed}
\usepackage{tablefootnote}
\usepackage{makecell}
\usepackage[normalem]{ulem}

\usepackage[colorlinks=true,pdfstartview=FitV, linkcolor=blue, citecolor=blue,urlcolor=blue]{hyperref} 
\usepackage[sort&compress,numbers,colon,merge]{natbib}

\makeatletter
\g@addto@macro\bfseries{\boldmath}
\makeatother

\usepackage[capitalise]{cleveref}
\usepackage{tikz} 
\usepackage{tkz-euclide}
\usetikzlibrary{backgrounds} 
\usetikzlibrary{decorations.pathmorphing}
\usetikzlibrary{arrows.meta}
\usetikzlibrary{shapes.misc}
\tikzset{
mystyle/.style={line width=1, baseline, scale=0.6, every node/.style={scale=1}},
v/.style={decorate, draw, decoration={snake, segment length=2.mm, amplitude=0.5mm}},
f/.style={draw, decoration={markings,mark=at position #1 with {\arrow[]{Latex[length=1.5mm,width=1.5mm]}}},
    postaction={decorate},node contents=#1},
f/.default=.6,
fb/.style={draw,decoration={markings,mark=at position #1 with {\arrowreversed[]{Latex[length=1.5mm,width=1.5mm]}}},
    postaction={decorate},node contents=#1},
fb/.default=.6,
s/.style={dashed,draw, postaction={decorate},
        decoration={markings,mark=at position .55 with {\arrow[very thick]{latex}}}},
sb/.style={dashed,draw, postaction={decorate},
        decoration={markings,mark=at position .35 with {\arrowreversed[draw=black,very thick]{latex}}}},
snar/.style={dashed,draw,line width =1.25pt},
cross/.style={cross out, draw=black, minimum size=2*(#1-\pgflinewidth), inner sep=0pt, outer sep=0pt}, 
         }
\usepackage{xcolor}
\definecolor{darkgreen}{rgb}{0,0.5,0}

\begin{document}
\begin{titlepage}
\vspace*{-1cm}
\phantom{hep-ph/***}
\flushright
\hfil{CPPC-2024-04}


\vskip 1.5cm
\begin{center}
\mathversion{bold}
{\LARGE\bf
Scalar dark matter explanation of the excess\\[2ex] in the Belle II $B^+\to K^+ +$ invisible measurement
}\\[3mm]
\mathversion{normal}
\vskip .3cm
\end{center}
\vskip 0.5  cm
\begin{center}
{\large Xiao-Gang He}$^{1}$,
{\large Xiao-Dong Ma}$^{2,3}$,
{\large Michael A.~Schmidt}$^{4}$,
{\large German Valencia}$^{5}$ and\\
{\large Raymond R. Volkas}$^{6}$
\\
\vskip .7cm
{\footnotesize
$^{1}$ Tsung-Dao Lee Institute, and School of Physics and Astronomy,
Shanghai Jiao Tong University, Shanghai 200240, China\\[0.3cm]
$^{2}$ Key Laboratory of Atomic and Subatomic Structure and Quantum Control (MOE), 
Guangdong Basic Research Center of Excellence for Structure and Fundamental Interactions of Matter, 
Institute of Quantum Matter, South China Normal University, Guangzhou 510006, China\\[0.3cm]
$^{3}$ Guangdong-Hong Kong Joint Laboratory of Quantum Matter, 
Guangdong Provincial Key Laboratory of Nuclear Science, 
Southern Nuclear Science Computing Center, South China Normal University, Guangzhou 510006, China\\[0.3cm]
$^{4}$ Sydney Consortium for Particle Physics and Cosmology, School of Physics, The University of New South Wales, Sydney, NSW 2052, Australia\\[0.3cm]
$^{5}$ School of Physics and Astronomy, Monash University,
Wellington Road, Clayton, Victoria 3800, Australia\\[0.3cm]
$^{6}$ ARC Centre of Excellence for Dark Matter Particle Physics,
School of Physics, The University of Melbourne, Victoria 3010, Australia\\[0.3cm]
\vskip .5cm
\begin{minipage}[l]{.9\textwidth}
\begin{center}
\textit{E-mail:}
\tt{hexg@sjtu.edu.cn},
\tt{maxid@scnu.edu.cn},
\tt{m.schmidt@unsw.edu.au},
\tt{german.valencia@monash.edu},
\tt{raymondv@unimelb.edu.au}
\end{center}
\end{minipage}
}
\end{center}
\vskip 1cm
\begin{abstract}

Recently Belle II reported the first measurement of $B^+\to K^++{\rm invisible (inv)}$, which is $2.7\sigma$ above the standard model (SM) prediction. If confirmed, this calls for new physics beyond SM. In the SM, the invisible particles are neutrino-anti-neutrino pairs. There are more possibilities when going beyond the SM. 
In this work, we focus on decays to dark matter (DM) and show that the $B\to K +\mathrm{inv}$ excess from Belle II and DM relic density can be simultaneously explained in a simple extension of the SM. The model introduces a real scalar singlet $\phi$ acting as a DM candidate, and two heavy vector-like quarks $Q,D$ with the same quantum numbers as the SM left-handed quark doublet and right-handed down-type quark singlet, respectively. All these new particles are odd under a $\mathbb{Z}_2$ symmetry while the SM particles are even. The model can successfully explain the Belle II anomaly and DM relic density for TeV-scale heavy quarks with hierarchical Yukawa couplings involving $b$ and $s$ quarks. At the same time, it can easily satisfy other flavour physics constraints. Direct detection searches utilizing the Migdal effect constrain some of the parameter space.
\end{abstract}
\end{titlepage}


\section{Introduction}

The $B\to K^{(*)} \nu \bar \nu$ decays are among the cleanest modes to search for new physics,
due to their well-controlled theoretical uncertainty in the standard model (SM). 
The Belle II experiment recently measured the charged decay mode $B^+ \to K^+ \nu\bar\nu$  \cite{Belle-II:2023esi} 
\begin{align}
\label{eq:newm}
{\cal B}(B^+ \to K^+ \nu \bar \nu)_{\tt exp}=(2.3 \pm 0.7) \times 10^{- 5},
\end{align}
which is $2.7\,\sigma$ in excess of the SM prediction~\cite{Becirevic:2023aov},
\begin{align}
{\cal B}(B^+\to K^+\nu\bar\nu)_{\tt SM} = (4.43\pm 0.31)\times 10^{-6}\;,
\end{align}
where the tree-level contribution mediated by $\tau$ leptons has been subtracted. 
This result is also about twice as large, but consistent within errors,
with the previous weighted average 
${\cal B}(B^+\to K^+\nu\bar\nu)_{\tt exp}^{2021}= (1.1\pm0.4)\times 10^{-5}$ performed by the Belle II collaboration~\cite{Belle-II:2021rof},
and with the new weighted average 
${\cal B}(B^+\to K^+\nu\bar\nu)_{\tt exp}^{\tt ave}= (1.3\pm0.4)\times 10^{-5}$ \cite{Belle-II:2023esi} 
including the earlier results from Belle~\cite{Belle:2013tnz,Belle:2017oht} and BaBar~\cite{BaBar:2010oqg,BaBar:2013npw}.

The Belle II measurement (new average) suggests an excess that is 3.6--7 (2--4) times the SM prediction, and has stimulated theoretical exploration of viable new physics scenarios. 
These include new contributions to $B^+ \to K^+ \nu \bar\nu$
from heavy mediators within the SM effective field theory (SMEFT) framework or renormalizable models
\cite{Bause:2023mfe,Athron:2023hmz,Allwicher:2023xba,Amhis:2023mpj,He:2023bnk,Chen:2023wpb,Fridell:2023ssf,Chen:2024jlj,Loparco:2024olo,Hou:2024vyw,Chen:2024cll},
and new decay modes involving new light invisible particles like 
a single scalar \cite{Berezhnoy:2023rxx,Abdughani:2023dlr,Datta:2023iln,Altmannshofer:2023hkn,McKeen:2023uzo,Ho:2024cwk}, 
dark photon \cite{Altmannshofer:2023hkn,Gabrielli:2024wys},
sterile neutrinos \cite{Felkl:2023ayn,Dreiner:2023cms}, 
or dark matter (DM) pairs \cite{He:2023bnk,McKeen:2023uzo,Ho:2024cwk}. See also earlier discussions~\cite{Browder:2021hbl,He:2021yoz,Felkl:2021uxi,He:2022ljo,Ovchynnikov:2023von,Asadi:2023ucx} of the slight excess indicated by the earlier weighted average~\cite{Belle-II:2021rof}.

For the light DM explanation, some of us identified in Ref.~\cite{He:2023bnk} that the dimension-5 (dim-5) quark scalar current operator,\footnote{The operator has been studied earlier in \cite{Bird:2004ts} as an example for using $B\to K + \mathrm{inv}$ as a probe for dark matter.}  
$\mathcal{O}_{d\phi}^{S,sb} = (\bar{s} b)\, \phi^2$,
in the low-energy effective theory ($\phi$LEFT) extended by a light real scalar DM $\phi$, provides a good fit to the data with a high scale for the Wilson coefficient (WC) of $|C^{S,sb}_{d\phi}|^{-1}\sim 1.5 \times 10^{7}\, \mathrm{GeV}$ for $m_\phi=1$ GeV. The field $\phi$ is odd with respect to a $\mathbb{Z}_2$ symmetry. The operator may be generated from the dim-6 operator in SMEFT extended by $\phi$ ($\phi$SMEFT), 
\begin{align}
\label{eq:O1}
\mathcal{O}_{qdH\phi^2} = (\bar{q}_{L} d_{R} H)\, \phi^2,
\end{align}
where $H$ is the SM Higgs doublet. The focus of this paper is to establish a viable renormalizable ultraviolet (UV) completion of this operator, which simultaneously explains the observed DM relic density through the thermal freeze-out mechanism.

To explain the excess with new decay modes involving a light DM pair, 
we refer to the difference between the measurement and the SM value as the ``NP window''.
Using the new Belle II measurement together with the SM prediction the NP window is
\begin{align}
\label{eq:np_window}
{\cal B}(B^+ \to K^+ +\mathrm{inv})_{\tt NP} 
= (1.86\pm 0.67)\times 10^{-5},
\end{align} 
assuming three-body decay kinematics with massless neutrinos,
which becomes $(0.86\pm 0.40)\times 10^{-5}$ if the new average experimental value is used.
Usually, the interaction mediating $B^+\to K^+ + \mathrm{inv}$ also leads to the transitions $B^0\to K^0+\mathrm{inv}$ and $B\to K^*+\mathrm{inv}$, which may pose strong constraints on the scenarios. 
Currently, the strongest experimental constraints on these modes are  \cite{Belle:2013tnz,Belle:2017oht}
\begin{subequations}
\begin{align}
{\cal B}(B^0 \to K^{0} \nu \bar \nu) &\leq 2.6  \times 10^{- 5} {\rm~(90\%~c.l.)},
\\
{\cal B}(B^+ \to K^{+*} \nu \bar \nu) &\leq 4.0  \times 10^{- 5} {\rm~(90\%~c.l.)},
\\
{\cal B}(B^0 \to K^{0*} \nu \bar \nu) &\leq 1.8  \times 10^{- 5} {\rm~(90\%~c.l.)},
\end{align}
\end{subequations}
assuming three-body decay kinematics with massless neutrinos, as predicted by the SM.
Subtracting their SM predictions, these lead to the corresponding NP bounds 
for the relevant modes with invisible particles to be \cite{He:2022ljo}: 
$2.3  \times 10^{- 5},\, 3.1  \times 10^{- 5}$ and $1.0  \times 10^{- 5} $, respectively.

The paper is organized as follows.
In \cref{sec:model} we introduce the renormalizable UV model which generates the operator in \cref{eq:O1}. Its implications for the $B$ meson decays $B\to K^{(*)} +\mathrm{inv}$ are discussed in \cref{sec:BKss} and dark matter in \cref{sec:DM}. We discuss constraints and the available parameter space explaining both the excess in $B^+\to K^+ +\mathrm{inv}$ and the DM relic abundance in \cref{sec:discussion} and conclude in \cref{sec:conclusions}. 

\section{The Model}
\label{sec:model}

In order to realize the UV completion of the operator in~\cref{eq:O1}, we introduce two heavy vector-like quarks $Q \sim(\mathbf{3},\mathbf{2},1/6)$ and $D \sim(\mathbf{3},\mathbf{1},-1/3)$ with masses $m_Q$ and $m_D$, respectively, in addition to the light scalar field $\phi\sim(\mathbf{1},\mathbf{1},0)$ to generate the Yukawa interactions $\bar q_{Li} Q_R\phi$ and $\bar d_{Ri} D_L \phi$ with $Q_R \equiv P_R Q, D_L \equiv P_L D$. 
Here we use lower case letters $q_L,d_R,u_R$ to represent the SM left-handed quark doublet, 
right-handed down-type quark and up-type quark singlets, respectively. 
The three numbers in parentheses indicate how they transform with respect to the SM gauge group $\rm SU(3)_{\rm c}\times SU(2)_{\rm L}\times U(1)_{\rm Y}$.
All exotic fields --- $\phi$, $Q_{L,R}$ and $D_{L,R}$ --- are required to be odd under an imposed $\mathbb{Z}_2$ symmetry that stabilises the DM particle $\phi$. 
Taking $m_{Q,D}>m_\phi +m_q$, the vector-like quarks $Q$ and $D$ are unstable 
and decay into $\phi$ and a SM quark of mass $m_q$. 
Because of the $\mathbb{Z}_2$ symmetry, the vector-like quarks do not mix with SM quarks, 
but they mix among themselves after electroweak symmetry breaking. 
The full new physics parts of the Lagrangian (kinetic, Yukawa and Higgs potential) are
\begin{subequations} 
\label{eq:modelL}
\begin{align}
\mathcal{L}_{\tt kinetic}^{\tt NP} & = 
\bar Q i\slashed{D} Q - m_Q \bar Q Q 
+ \bar D i\slashed{D} D - m_D \bar D D 
+ \frac12 \partial_\mu \phi \partial^\mu \phi - \frac12 m_\phi^2 \phi^2,
\\
\mathcal{L}_{\tt Yukawa}^{\tt NP} & = 
y_{q}^p \bar{q}_{Lp} Q_{R} \phi + y_{d}^p \bar{D}_L d_{Rp} \phi 
- y_1 \bar{Q}_L D_R H - y_2 \bar{Q}_R D_L H + {\rm h.c.}\;,    
\\
V_{\tt potential}^{\tt NP}  & = \frac14 \lambda_\phi \phi^4 + \frac12 \kappa\, \phi^2 H^\dagger H\;, 
\end{align}
\end{subequations}
where group theory indices are suppressed and the SM quark generations are denoted by indices $p,r$. The choice of a minus sign with Yukawa couplings $y_{1,2}$ is for later matching convenience.
$H\sim(\mathbf{1},\mathbf{2},1/2)$ denotes the electroweak Higgs doublet. 
Using the freedom to rephase the two heavy vector-like quark fields, we choose $y_1$ and one of the Yukawa couplings $y_q^p$ to be real without loss of generality. All other Yukawa couplings generally are complex.

The DM particle $\phi$ couples to the SM Higgs doublet via the Higgs portal interaction $\phi^2 H^\dagger H$, in addition to its coupling to SM quarks via the Yukawa interactions. The former is constrained by invisible Higgs decay with the branching ratio BR($h\to \mathrm{inv})<0.107$ at 95\% CL \cite{ATLAS:2023tkt} (see also \cite{ATLAS:2022yvh,CMS:2022qva}), which leads to $\kappa \lesssim 0.01$ for light $\phi$ with mass below tens of GeV. In the following phenomenological study, we neglect the Higgs portal interaction and consequently also its contribution to the DM mass. Taking it small compared to the bare mass term implies $\kappa \ll 2 m_\phi^2/v^2\approx 3\times10^{-5} (m_\phi/{\rm GeV})^2$ which is consistent with radiative corrections from fermion loops.

\section{\texorpdfstring{$B\to K^{(*)} + \mathrm{inv}$}{B->K+inv}}
\label{sec:BKss}

At energies  below the masses $m_Q\simeq m_D$ of the heavy vector-like quarks $D$ and $Q$, 
the interactions of the scalar field $\phi$ with SM quarks are described
to leading order by the following $\phi$SMEFT Lagrangian,
\begin{align}
\label{eq:SMEFT}
\mathcal{L}_{\phi\phi qq}^{\tt SMEFT} & = 
C_{qdH\phi^2}^{pr} \mathcal{O}_{qdH\phi^2}^{pr}
+C_{quH\phi^2}^{pr} \mathcal{O}_{quH\phi^2}^{pr}
+\mathrm{h.c.}\;,
\end{align}
with the first effective operator given in \cref{eq:O1} and the second, 
$\mathcal{O}_{quH\phi^2}^{pr}  = (\bar q_{Lp} u_{Rr} \tilde H) \phi^2$ with
$\tilde H = \epsilon H^*$.

\begin{figure}[tb!]
\centering
{\scriptsize
\begin{tikzpicture}[mystyle,scale=0.7]
\begin{scope}[shift={(1,1)}]
\draw[f] (-1.5,1.5) node[left]{$d_R$} -- (0,0);
\draw[f,blue] (0,0) -- (1.5,0) node[midway, yshift =  7 pt]{$D_L$};
\draw[f,blue] (1.5,0) -- (3,0) node[midway, yshift =  7 pt]{$D_R$};
\draw[f,blue] (3,0) -- (4.5,0) node[midway, yshift =  7 pt]{$Q_L$};
\draw[f,blue] (4.5,0) -- (6,0) node[midway, yshift =  7 pt]{$Q_R$};
\draw[f] (6,0) -- (7.5,1.5)node[right]{$q_L$};
\draw[sb] (3,0) -- (3,-2) node[midway, xshift = - 7 pt]{$H$};
\draw[snar, darkgreen] (0,0) -- (0,-2) node[midway, xshift = - 7 pt]{$\phi$};
\draw[snar, darkgreen] (6,0) -- (6,-2) node[midway, xshift = 7 pt]{$\phi$};
\draw (1.5, 0) node[cross=4pt, magenta] {};
\draw (4.5, 0) node[cross=4pt, magenta] {};
\end{scope}
\end{tikzpicture}
\begin{tikzpicture}[mystyle,scale=0.7]
\begin{scope}[shift={(1,1)}]
\draw[f] (-1.5,1.5) node[left]{$d_R$} -- (0,0);
\draw[f,blue] (0,0) -- (1.5,0) node[midway, yshift =  7 pt]{$D_L$};
\draw[f] (1.5,0) -- (3,0) node[midway, yshift =  7 pt]{$d_R$};
\draw[f] (3,0) -- (4.5,1.5)node[right]{$q_L$};
\draw[snar, darkgreen] (0,0) -- (0,-2) node[midway, xshift = - 7 pt]{$\phi$};
\draw[snar, darkgreen] (1.5,0) -- (1.5,-2) node[midway, xshift = - 7 pt]{$\phi$};
\draw[sb] (3,0) -- (3,-2) node[midway, xshift =  7 pt]{$H$};
\end{scope}
\end{tikzpicture}
\begin{tikzpicture}[mystyle,scale=0.7]
\begin{scope}[shift={(1,1)}]
\draw[f] (-1.5,1.5) node[left]{$d_R$} -- (0,0);
\draw[f] (0,0) -- (1.5,0) node[midway, yshift =  7 pt]{$q_L$};
\draw[f,blue] (1.5,0) -- (3,0) node[midway, yshift =  7 pt]{$Q_R$};
\draw[f] (3,0) -- (4.5,1.5)node[right]{$q_L$};
\draw[snar, darkgreen] (3,0) -- (3,-2) node[midway, xshift =  7 pt]{$\phi$};
\draw[snar, darkgreen] (1.5,0) -- (1.5,-2) node[midway, xshift = 7 pt]{$\phi$};
\draw[sb] (0,0) -- (0,-2) node[midway, xshift = - 7 pt]{$H$};
\end{scope}
\end{tikzpicture}
\begin{tikzpicture}[mystyle,scale=0.7]
\begin{scope}[shift={(1,1)}]
\draw[f] (-1.5,1.5) node[left]{$d_R$} -- (0,0);
\draw[f,blue] (0,0) -- (1.5,0) node[midway, yshift =  7 pt]{$D_L$};
\draw[f,blue] (1.5,0) -- (3,0) node[midway, yshift =  7 pt]{$Q_R$};
\draw[f] (3,0) -- (4.5,1.5)node[right]{$q_L$};
\draw[sb] (1.5,0) -- (1.5,-2) node[midway, xshift = - 7 pt]{$H$};
\draw[snar, darkgreen] (0,0) -- (0,-2) node[midway, xshift = - 7 pt]{$\phi$};
\draw[snar, darkgreen] (3,0) -- (3,-2) node[midway, xshift = 7 pt]{$\phi$};
\end{scope}
\end{tikzpicture}
}%
\caption{Feynman diagrams contributing to the matching to the $\phi$SMEFT-like operator $\mathcal{O}_{qdH\phi^2}$ via $t$-channel exchange of the vector-like fermions $Q$ and $D$. The magenta crosses represent mass insertions.}
\label{fig:tchannel}
\end{figure}
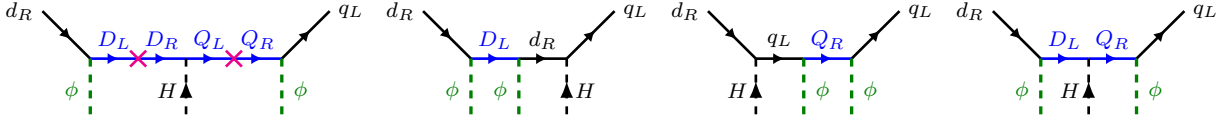

The WCs $C_{qdH\phi^2}$ and $C_{quH\phi^2}$ can be determined by integrating out
the heavy vector-like quarks $Q$ and $D$ at tree level as shown in~\cref{fig:tchannel}. 
With the aid of the \texttt{Matchete} code \cite{Fuentes-Martin:2022jrf}, we obtain the matching results,\footnote{The matching proceeds through an intermediate step that uses the two redundant operators, 
$(\bar q_{Lp} i\slashed{D} q_{Lr}) \phi^2$ 
and $(\bar d_{Rp} i\slashed{D} d_{Rr}) \phi^2$.
These are related to the operators in \cref{eq:SMEFT} via the equations of motion of the SM quark fields $q_L$ and $d_R$.
}
\begin{subequations}
 \begin{align}
C_{qdH\phi^2}^{pr} & = \frac{ y_q^p y_d^r y_1}{m_Q m_D }
+ \frac{ y_q^p y_q^{x*} (Y_d)_{xr}}{2 m_Q^2}
+ \frac{ (Y_d)_{px} y_d^{x*} y_d^r}{2 m_D^2},   
\\
C_{quH\phi^2}^{pr} & = \frac{ y_q^p y_q^{x*} (Y_u)_{xr}}{2 m_Q^2},
\end{align}
\end{subequations}
where $Y_u$ and $Y_d$ denote the SM up-quark and down-quark Yukawa couplings, 
$\bar q_L Y_u u_R \tilde H + \bar q_L Y_d d_R H$, respectively. 
The first three Feynman diagrams in \cref{fig:tchannel} correspond to the three terms in $C_{qdH\phi^2}^{pr}$,
the fourth diagram yields  dim-8 operators that are expected to be suppressed by two additional heavy mass powers and thus will be neglected. 
The coefficient $C_{quH\phi^2}^{pr}$ is generated by a similar diagram to the third one in \cref{fig:tchannel} in which the $d_R$ and $H$ fields are replaced by $u_R$ and $\tilde H$ respectively. 

To study $B \to K^{(*)}+\mathrm{inv}$ and other low energy processes much lower than the EW scale, 
we need to match the $\phi$SMEFT to the $\phi$LEFT by integrating out the heavy SM states (top, Higgs, $W$ and $Z$) after EW symmetry breaking. 
Taking the down-type quark flavour eigenstates to be mass eigenstates, we obtain four $\phi$LEFT operators
\begin{align}
\label{eq:LEFT}
\mathcal{L}_{\phi\phi qq}^{\tt LEFT} =
\frac12 C_{d\phi}^{S,ij} (\bar d_i d_j) \phi^2 
+ \frac12 C_{d\phi}^{P,ij} (\bar d_i i\gamma_5 d_j) \phi^2 
+ \frac1 2 C_{u\phi}^{S,ij} (\bar u_i u_j) \phi^2 
+ \frac1 2 C_{u\phi}^{P,ij} (\bar u_i i\gamma_5 u_j) \phi^2,     
\end{align}
with their corresponding WCs
\begin{subequations}
\label{eq:leftWCmatching}
\begin{align}
C_{d\phi}^{S,ij} & =
 \frac{ (y_q^i y_d^j y_1 + y_q^{j*} y_d^{i*} y_1^* )v}{\sqrt{2}m_Q m_D }
+ \left( \frac{ y_q^i y_q^{j*}}{2 m_Q^2}
+ \frac{  y_d^{i*} y_d^j}{2 m_D^2} \right)  (m_{d_i} +m_{d_j}), 
\\
i C_{d\phi}^{P,ij} & =
 \frac{ (y_q^i y_d^j y_1 - y_q^{j*} y_d^{i*} y_1^*)v}{\sqrt{2}m_Q m_D }
- \left( \frac{ y_q^i y_q^{j*}}{ 2 m_Q^2}
- \frac{  y_d^{i*} y_d^j}{2 m_D^2} \right)  (m_{d_i} - m_{d_j}), 
\\
C_{u\phi}^{S,ij} & = \frac{ \tilde y_q^i \tilde y_q^{j*}}{2 m_Q^2}(m_{u_i} +m_{u_j}),
\\
i C_{u\phi}^{P,ij} & =  - \frac{ \tilde y_q^i \tilde y_q^{j*}}{2 m_Q^2} (m_{u_i} - m_{u_j})\;,
\end{align}
\end{subequations}
where $\tilde y_q^i \equiv (V_{\tt CKM})_{i x} y_q^x$ with $V_{\tt CKM}$ being the CKM matrix. 
In the above we use $i,j$ to represent the quark flavours in the mass basis.
Note that there are no vector operators of the form $(\bar q \gamma^\mu q) (\phi\overleftrightarrow{\partial_\mu} \phi)$ for a real scalar $\phi$ due to $(\phi\overleftrightarrow{\partial_\mu} \phi)=0$. 

The $B\to K\phi\phi$ ($B\to K^*\phi\phi$) transition is induced by the operator 
$\bar s b\phi^2$ ($\bar s i\gamma_5 b \phi^2$).
Their differential branching ratios have been calculated in \cite{He:2022ljo}\footnote{Note that a factor of $1/2$ needs to be included to account for the double counting of phase space integration for the identical $\phi$ pair. }
and are given by 
\begin{subequations}
\label{eq:B2Kphi2dis}
\begin{align}
\frac{d\Gamma_{B\to K\phi\phi}}{ds_B} & = \frac{\left|C_{d\phi}^{S,sb}\right|^2  m_B}{512 \pi^3}
f_0^2\frac{(1-x_K)^2 \lambda^{\tfrac12}(1,x_K,s_B) \sqrt{1- 4 x_\phi/x_B}}{(\sqrt{x_b}-\sqrt{x_s})^2}, 
\\
\frac{d\Gamma_{B\to K^*\phi\phi}}{ds_B} & = \frac{\left|C_{d\phi}^{P,sb}\right|^2 m_B}{512 \pi^3} A_0^2 \frac{\lambda^{\tfrac32}(1,x_{K^*},s_B) \sqrt{1- 4 x_\phi/x_B} }{(\sqrt{x_b}+\sqrt{x_s})^2},
\end{align}
\end{subequations}
in terms of the K\"all\'en function $\lambda(x,y,z)=x^2+y^2+z^2-2xy-2xz-2yz$. 
All dimensionful quantities are normalized relative to the $B$ meson mass: 
the squared invariant mass of the DM pair is $q^2 = s_B m_B^2$ 
and the mass ratios are 
$x_{K^{(*)}}=m_{K^{(*)}}^2/m_B^2$, $x_\phi=m_\phi^2/m_B^2$, 
$x_b=m_b^2/m_B^2$, 
and $x_s=m_s^2/m_B^2$. 
The $q^2$ dependence of the hadronic form factors $f_0$ and $A_0$ associated with scalar and pseudo-scalar quark currents is left implicit for notational simplicity, and we use the recent fitting results from \cite{Gubernari:2023puw} for the later numerical analysis. 

\begin{figure}
\centering
\includegraphics[width=0.7\linewidth]{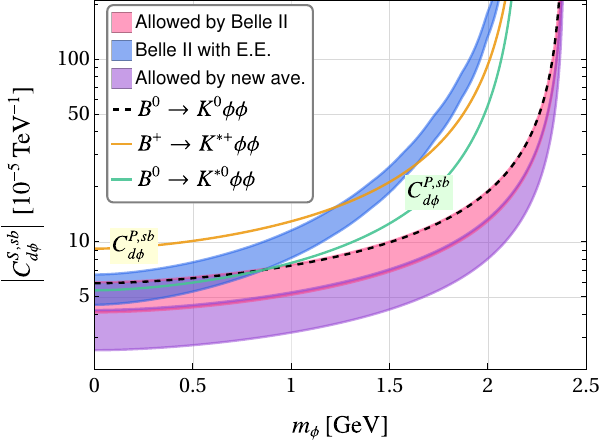}
\caption{Preferred parameter space to explain the excess in $B^+\to K^+ +\mathrm{inv}$ via additional decay channels to DM final states, $B^+\to K^+\phi\phi$ based on the latest Belle II measurement~\cite{Belle-II:2023esi} in red, with an additional reweighted one to account for the selection efficiency in blue, and finally based on the new average in purple~\cite{Belle-II:2023esi}. The dashed line indicates the current constraint from $B^0\to K^0+\mathrm{inv}$~\cite{Belle:2013tnz,Belle:2017oht} on $|C^{S,sb}_{d\phi}|$ and the solid orange and green lines the constraints on $|C^{P,sb}_{d\phi}|$ posed by the searches for $B^+\to K^{+*}+\mathrm{inv}$ and $B^0\to K^{0*}+\mathrm{inv}$~\cite{Belle:2013tnz,Belle:2017oht}, respectively.
}
\label{fig:btokss}
\end{figure}

From \cref{eq:B2Kphi2dis} it can be seen that the differential branching ratios with final $K$ and $K^*$ states depend on different $\phi$LEFT WCs. 
Hence, it is possible to obtain a large enhancement in $B\to K\phi\phi$ without substantially modifying $B\to K^*\phi\phi$. 
In fact for $y_{q}^i = y_{d}^{i*}$ there is a cancellation in $C_{d\phi}^{P,ij}$ and thus there are no additional contributions to $B\to K^*\phi\phi$. 
In terms of the $\phi$LEFT WCs $C_{d\phi}^{S,sb}$, we show the allowed parameter space to accommodate the Belle II excess in \cref{eq:np_window} (the new average number) as the magenta (purple) region in \cref{fig:btokss}. It can be seen the excess in $B^+\to K^++ \mathrm{inv}$ can be explained for $C^{S,sb}_{d\phi}\sim (3-8)/(10^5\, \rm TeV)$ for a DM mass of $m_\phi=1$ GeV, without conflicting with the bound from the neutral kaon mode shown as the black dashed line.
The same plot also shows the constraints on the $C_{d\phi}^{P,sb}$ from the vector kaon final states as orange and blue curves, since they are also generated in the model. As can be seen, the neutral mode
puts more a stringent bound on $C_{d\phi}^{P,sb}$ than the charged mode. 

The situation depicted in \cref{eq:np_window} would be modified by experimental efficiency (E.E.) considerations. 
The inclusion of the Belle II signal selection efficiency modifies the allowed parameter space. To illustrate the impact this effect has on our study we include the blue region in the figure. This region is obtained by scaling the predictions of our model by a ratio of normalized rates weighted by the reported efficiency $\epsilon$,
\begin{align}
\omega (m)
= \frac{\sum_i   \tilde \Gamma_{i, \rm SM} \epsilon_i }{\sum_i  \tilde \Gamma_{i, \rm NP}(m) \epsilon_i }. 
\end{align}
In this factor, $\tilde \Gamma_{i, \rm NP}(m)$ ($\tilde \Gamma_{i, \rm SM}$) is the normalized width from the new physics (SM) contribution in the $i$-th bin \cite{He:2022ljo}, and $\epsilon_i$ is the experimental efficiency for that bin in the inclusive tag analysis reported in \cite{Belle-II:2023esi}. 
It can be seen that the consideration of efficiency requires a slightly larger $C_{d\phi}^{S,sb}$ for each fixed DM mass benchmark. Note that the constraints from other $B\to K^{(*)} + \mathrm{inv}$ decays may also be relaxed due to experimental efficiency considerations, which depend on the experimental analysis. See Ref.~\cite{Gartner:2024muk} for a detailed discussion on how the experimental efficiencies could be incorporated. In \cref{sec:discussion}, we combine these results with that of DM analysis to identify the viable parameter space.

\section{Dark matter}
\label{sec:DM}

As invisible Higgs decay constrains the Higgs portal coupling to be small and thus DM annihilations via the Higgs portal $\phi \phi \to SM$ to be subdominant, we focus on thermal freeze-out via DM annihilation generated through $t$-channel exchange of the heavy vector-like quarks, which is described by the effective operators in \cref{eq:LEFT}. The explanation of the excess $B^+\to K^+\phi\phi$ prefers light dark matter, and thus we use chiral perturbation theory to evaluate the DM annihilation cross section into light mesons. Taking into account the three light quarks, the QCD Lagrangian extended by the DM-quark interactions in \cref{eq:LEFT} takes the form 
\begin{align}
\label{eq:exter}
{\cal L} \ni  \mathcal{L}_{\mathrm{QCD}} 
-\left[\overline{q_R}(s+i p) q_L
+\text { h.c. }\right]\;,
\end{align} 
where the scalar and pseudo-scalar sources are 
\begin{align}
   s = - \frac12
   \begin{pmatrix}
     C_{u\phi}^{S,uu}  & 0 & 0 \\
     0 & C_{d\phi}^{S,dd} & C_{d\phi}^{S,ds} \\
     0 & C_{d\phi}^{S,sd} & C_{d\phi}^{S,ss} 
   \end{pmatrix}
   \phi^2, \quad 
   p = \frac12
   \begin{pmatrix}
     C_{u\phi}^{P,uu}  & 0 & 0 \\
     0 & C_{d\phi}^{P,dd} & C_{d\phi}^{P,ds} \\
     0 & C_{d\phi}^{P,sd} & C_{d\phi}^{P,ss} 
   \end{pmatrix}
   \phi^2
   \;.
\end{align}
The leading order DM-meson interactions appears at ${\cal O}(p^2)$ in chiral power counting \cite{Gasser:1983yg,Gasser:1984gg}, 
and thus the relevant part of the effective Lagrangian in chiral perturbation theory is 
\begin{align}
\label{eq:p2l}
\mathcal{L}_{p^2}
= 
\frac{F_0^2}{4} {\rm tr}\left[(D_\mu U) (D^\mu U)^\dagger \right] + 
\frac{F_0^2}{4}{\rm tr} \left[\chi U^\dagger +U\chi^\dagger \right],
\end{align}
where $F_0\approx 87\,\rm MeV$ is the pion decay constant in the chiral limit, $\chi = 2 B(s-ip)$ with $B\approx 2.8\,\rm GeV$,
and $U$ is related to the mesons by
\begin{align}
U=\text{exp}\left[i\frac{\sqrt{2} \Phi}{F_0}\right], \quad
\Phi & = 
\begin{pmatrix}
\frac{\pi^0}{\sqrt{2}}+\frac{\eta}{\sqrt{6}} & \pi^+ & K^+
\\
\pi^- & -\frac{\pi^0}{\sqrt{2}}+\frac{\eta}{\sqrt{6}} & K^0
\\
K^- & \bar{K}^0 & -\sqrt{\frac{2}{3}}\eta
\end{pmatrix}. 
\end{align}
Expanding $U$ to second order in the meson field $\Phi$,
the leading order Lagrangian for DM-meson interactions reads
\begin{align}
  \mathcal{L}_{\phi P} \ni - \sqrt{2} B F_0 \, {\rm tr}\left[ p \Phi\right]
  - B \, {\rm tr}\left[ s \Phi^2 \right]
  + {\cal O}(\phi^2 P^3)
  \;.
\end{align}
In the limit $y_1 v \gg m_{u,d,s}$ and $y_{q}^i\simeq y_{d}^{i*}$, which is preferred by the explanation of the excess in $B^+\to K^++\mathrm{inv}$, 
the leading contributions originate from the second term and result in 
\begin{align}
   \mathcal{L}_{\phi P} 
   & \ni \frac{ B}{2} \phi^2
   \left\{
   \left(C_{u\phi}^{S,uu}+ C_{d\phi}^{S,dd}\right) \left( \pi^+ \pi^- + \frac12 \pi^0 \pi^0 \right)
  + \left(C_{u\phi}^{S,uu}+ C_{d\phi}^{S,ss}\right) K^+ K^-
   \right.
   \nonumber
   \\
   & 
   + (C_{d\phi}^{S,dd}+ C_{d\phi}^{S,ss}) K^0 \bar K^0
   + \frac{1}{6} \left( C_{u\phi}^{S,uu}+ C_{d\phi}^{S,dd} +  4 C_{d\phi}^{S,ss} \right)\eta^2 
+ \frac{1}{\sqrt{3}}\left( C_{u\phi}^{S,uu} -  C_{d\phi}^{S,dd} \right) \pi^0 \eta 
\nonumber
   \\
   & \left. + \left[ 
   C_{d\phi}^{S,ds}
   \left( 
    \pi^+ K^- 
    - \frac{1}{\sqrt{2}} \pi^0 \bar K^0
    - \frac{1}{\sqrt{6}} \eta \bar K^0
   \right)
   + {\rm h.c.}
   \right ] \right\} \;.
\end{align}
The WC $C_{u\phi}^{S,uu}$ can be neglected because it is suppressed by small CKM matrix elements in addition to $m_u\ll y_1 v$.
The down quark Yukawa coupling $y_{q,d}^d$ is not necessary for  $B\to K^{(*)} \phi\phi$ nor for the DM annihilation rate. However, it becomes a consideration when we look at direct detection bounds that rely on the Migdal effect. For illustration, we will use a value $|y_{q,d}^d|\sim 0.2 |y_{q,d}^s|$, leading to $|C_{d\phi}^{S,dd}|\sim 0.2 |C_{d\phi}^{S,ds}|\sim 0.04 |C_{d\phi}^{S,ss}|$,  and thus the dominant annihilation channels are into pairs of kaons and $\eta$ mesons,
\begin{subequations}
\begin{align}
     \sigma(\phi\phi\to K^+ K^-, K^0 \bar K^0) 
     & = \frac{B^2|C_{d\phi}^{S,ss}|^2}{16\pi s} \left(\frac{s-4 m_K^2}{s-4m_\phi^2}\right)^{1/2}
     \;,
     \\
     \sigma(\phi\phi\to \eta\eta) 
     & = \frac{B^2 |C_{d\phi}^{S,ss}|^2}{18\pi s} \left(\frac{s-4 m_\eta^2}{s-4m_\phi^2}\right)^{1/2}
     \;.
\end{align}
\end{subequations}
The thermal average takes the form \cite{Gondolo:1990dk}
\begin{align}\label{eq:thermal_xs}
  \langle \sigma v \rangle  
  = \frac{4 x}{K_2^2(x)} \int_0^\infty d \epsilon \, \epsilon \sqrt{1+\epsilon} K_1(2 x\sqrt{1+\epsilon}) \sigma, \quad 
   \epsilon \equiv {s - 4 m_\phi^2 \over 4 m_\phi^2},
\end{align}
where $K_{1,2}$ are modified Bessel functions and $x \equiv m_\phi/T$ with $T$ being temperature. 
Thus the thermally averaged cross sections for DM annihilation into kaons and $\eta$ mesons are
\begin{subequations}
\begin{align}
    \langle \sigma v (\phi\phi\to K^+K^-,K^0\bar K^0)\rangle 
    & = \frac{B^2 |C_{d\phi}^{S,ss}|^2 \eta(x,z_K)}{64\pi m_\phi^2}  
    \;,
    \\
    \langle \sigma v (\phi\phi\to \eta\eta)\rangle 
    & = \frac{B^2 |C_{d\phi}^{S,ss}|^2 \eta(x,z_\eta)}{72\pi m_\phi^2},  
\end{align}
\end{subequations}
in terms of $z_{K,\eta}=m_{K,\eta}^2/m_\phi^2$ and the function 
\begin{align}
 \eta(x, z) & \equiv {4 x\over K_2^2(x)} \int_0^\infty d \epsilon \, {\sqrt{\epsilon} \sqrt{1 + \epsilon - z} \over \sqrt{1 +\epsilon} }  K_1(2 x\sqrt{1+\epsilon}) = {\cal O}(1) \;,
\end{align}
which takes values between 0.5 and 1.7 for the relevant parameter space. 
The relic density of WIMP dark matter is given by 
\begin{equation} 
\Omega_\phi = \frac{m_\phi\, n_{\phi,\text{fo}}\, s_0/s_{\text{fo}}}{\rho_{\rm crit}}
\;,
\end{equation}
where $n_{\phi,\text{fo}}$ denotes the dark matter number density at freeze-out,  $s_{0,\text{fo}}$ the entropy density today and at freeze-out respectively, and $\rho_{\rm crit}$ the critical energy density. 
The required thermally averaged DM annihilation cross section is approximately given by~\cite{Steigman:2012nb}
$\langle\sigma v \rangle \simeq 2.4 \times 10^{-26}\, \frac{\mathrm{cm}^3 \mathrm{s}^{-1}}{(\hbar c)^2 c} = 2.2\times 10^{-9}\, \mathrm{GeV}^{-2}$,
neglecting the slight variation in the required annihilation cross section between the kaon mass $m_K$ and $1$ GeV. A more precise prediction can be obtained using the results from 
Figures 4 and 5 in Ref.~\cite{Steigman:2012nb}. We digitise those two figures to obtain the matching point $x_*$ of the two regimes of the freeze-out process as a function of the DM mass and the thermally averaged cross section at $x_*$ which is required to obtain the correct relic abundance, and use it to predict the required value of the WC $|C_{d\phi}^{S,ss}|$.

\begin{figure}
\centering
\includegraphics[width=0.7\linewidth]{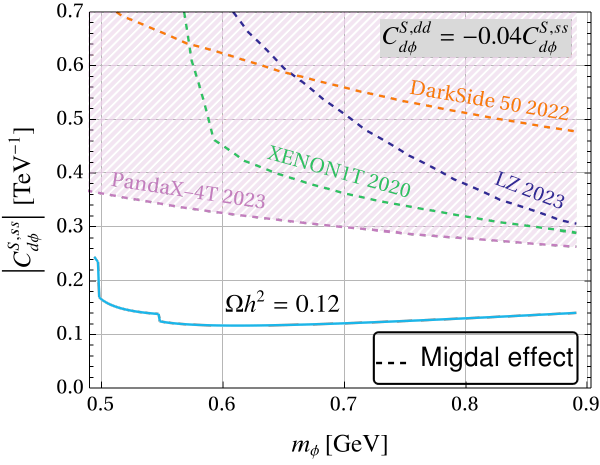}
\caption{The Wilson coefficient $|C^{S,ss}_{d\phi}|$, for which the observed DM relic abundance is obtained through annihilation into kaons and $\eta$ mesons, as a function of the DM mass $m_\phi$. For the benchmark point $R_{d/s}=-0.04$, the hatched region is excluded by the DM direct detection experiments via the Migdal effect. }
\label{fig:dmabundance}
\end{figure}

While annihilations to final states with more than two mesons are suppressed, we would have to include annihilations to $K^*$ meson pairs were this channel available. However, we restrict ourselves to DM masses below the $K^*$ threshold, which are also favoured by the explanation of the excess in $B^+\to K^+ +\mathrm{inv}$. In \cref{fig:dmabundance} we present the WC $|C_{d\phi}^{S,ss}|$ required to produce the correct DM relic density as a function of the DM mass $m_\phi$.
A comparison with the required value of $C_{d\phi}^{S,sb}\sim (3-8)/(10^5\,{\rm TeV})$ for explaining the Belle II anomaly clearly demonstrates the need for a large ($\sim 10^3$) hierarchy between the strange and bottom couplings.

\pmb{Direct detection}: 
Finally, we comment on the constraints from direct and indirect detection. 
Since the sensitivity of direct detection decreases significantly for sub-GeV scale DM
due to detection threshold issues, direct detection experiments using the usual nuclear recoil signal do not provide any meaningful constraints for the parameter space of interest.
However,  the Migdal effect can be used to bypass this kinematic restriction and probe the low mass DM region \cite{Migdal:1941,Ibe:2017yqa,Dolan:2017xbu}. The liquid xenon and argon experiments including XENON1T \cite{XENON:2019zpr}, DarkSide50 \cite{DarkSide:2022dhx}, LZ
 \cite{LZ:2023poo}, and PandaX-4T \cite{PandaX:2023xgl} have already used this technique to constrain the light DM coupling to nucleons, resulting in meaningful constraints on the spin-independent DM-nucleon cross section. To translate the experimental bound into constraints on our parameter space, we need to calculate the corresponding DM-nucleon cross section in the limit of zero momentum transfer within our scenario. This is done in the following way. 
 For the non-vanishing WCs $C_{d\phi}^{S,ss}$ and $C_{d\phi}^{S,dd}$, the corresponding operators 
${\cal O}_{d\phi}^{S,ss}={1\over 2} \bar s s \phi^2$ and ${\cal O}_{d\phi}^{S,dd}={1\over 2} \bar d d\phi^2$ can induce the usual spin-independent DM-nucleon non-relativistic operator ${\cal O}_1^N \equiv 1_\phi 1_N$ ($N=p/n$ for proton/neutron) with the coefficient,\footnote{See, for instance, Table 5 in \cite{DelNobile:2021wmp}.} 
\begin{align}
 c_1^N =  2 {m_N^2 \over m_s} f_{T_s}^{(N)} C_{d\phi}^{S,ss} + 2 {m_N^2 \over m_d} f_{T_d}^{(N)} C_{d\phi}^{S,dd}
 =  2 {m_N^2 \over m_s} f_{T_s}^{(N)} C_{d\phi}^{S,ss}\left( 1  + {m_s f_{T_d}^{(N)} \over m_d f_{T_s}^{(N)} } R_{d/s}\right),
\end{align}
where in the second step we have introduced $R_{d/s}\equiv { C_{d\phi}^{S,dd}/C_{d\phi}^{S,ss} }$. The masses of strange and down quarks are denoted $m_{s,d}$, and $f_{T_{s,d}}^{(N)}$ are nucleon form factors due to the scalar quark current at zero momentum transfer.  
For our numerical estimates, we use $m_s=93.4\,\rm MeV$, $m_s/m_d\approx 19.5$, $f_{T_s}^{(N)}=0.044$, $f_{T_d}^{(p)}=0.038$, and $f_{T_d}^{(n)}=0.056$  \cite{DelNobile:2021wmp}. 
A nonzero value of  $R_{d/s}$, $c_1^p\neq c_1^n$, corresponds to the isospin-violating DM scenario. 

Taking into account these isospin-violating couplings, the corresponding effective DM-nucleon cross section is calculated to be (see Eq.\,(H.37) in \cite{DelNobile:2021wmp}),
\begin{align}
\sigma_{\phi N} & = {\mu_{\phi N}^2  \over 4\pi m_\phi^2} \left| {m_N\over m_s} f_{T_s}^{(N)}{\cal C}_{d\phi}^{S,ss} \right|^2
\left| 
 \left( 1 + {m_s \over m_d} { f_{T_d}^{(p)} \over f_{T_s}^{(p)} } R_{d/s}\right) {Z \over A} 
 + 
\left( 1 + {m_s \over m_d} { f_{T_d}^{(n)} \over f_{T_s}^{(n)} }R_{d/s} \right){A-Z\over A}
\right|^2,  
\nonumber
\\
& \approx {\mu_{\phi N}^2 \over 4\pi m_\phi^2} \left| {m_N \over m_s} f_{T_s}^{(N)}{\cal C}_{d\phi}^{S,ss} \right|^2
\left| 
 \left( 1 + 16.84 R_{d/s}\right) {Z \over A} 
 + 
\left( 1 + 24.82 R_{d/s} \right){A-Z\over A}
\right|^2,  
\end{align}
where $\mu_{\phi N}$ is the reduced mass of the DM-nucleon system. 
In the second line we have used the numerical values given above for the hadronic parameters.  The DarkSide experiment is a pure argon detector with only one isotope $\rm{}^{40}Ar$. For the xenon target experiments (XENON1T, LZ, PandaX-4T), several isotopes are involved with  abundances summarised in Table.\,1 in \cite{DelNobile:2021wmp}. We 
need to use the weighted average $\sum \xi_i \sigma_{\phi N}^i$, where the sum runs over all isotopes $i$ with abundance $\xi_i$.  
To obtain the Midgal effect sensitivity curve in the $m_\phi-C_{d\phi}^{S,ss}$ plane, we fix 
$R_{d/s} = -0.04$, and show the constraints from different experiments as dashed lines in \cref{fig:dmabundance}.\footnote{Using the statistical procedures outlined in \cite{Liang:2024tef}, we find the XENON1T constraint is weakened by approximately a factor of 2.} As can be seen, the current constraints from the Migdal effect can already probe a large portion of the parameter space. In particular, the regions with  $R_{d/s} \gtrsim -0.02$ or $R_{d/s} \lesssim -0.07$, are already excluded. The remaining region for $R_{d/s}$ ($-0.07\lesssim R_{d/s} \lesssim -0.02$) can be further constrained by future DM direct detection experiments. 

\pmb{Indirect detection}:
For indirect detection, the main annihilation channel is into photon pairs. 
The DM-photon interactions arise at the one-loop order and take the form, 
${\cal O}_{\phi \gamma} = \phi^2 F_{\mu\nu} F^{\mu\nu}$, 
with the corresponding WC given by
\begin{align}
C_{\phi \gamma } 
= \frac{\alpha_{\tt em}}{36\pi}\left( \frac{5 y_q^x y_q^{*x}}{m_Q^2}
+ \frac{ y_d^x y_d^{*x}}{m_D^2}\right).  
\end{align}
The velocity-averaged DM annihilation cross section for $\phi\phi \to \gamma\gamma $ is estimated to be 
$\langle \sigma v \rangle_{\phi\phi \to \gamma\gamma} \approx (2/\pi)m_\phi^2|C_\phi \gamma|^2$ in the $s$-wave approximation. The current constraint on $\langle \sigma v (\phi\phi \to \gamma\gamma)\rangle$ of around ${\cal O}(10^{-28}\, \rm cm^3/s)$ \cite{Boddy:2015efa}, leads to 
$m_\phi |C_{\phi \gamma }| \lesssim 0.004/\rm TeV$. For $m_\phi \sim 0.5\,\rm GeV$ and ${\cal O}(1)$ Yukawa couplings, the constraint implies $m_Q \gtrsim 10\,\rm GeV$, a quite weak bound that can be easily satisfied. 
On the other hand, ${\cal O}_{\phi\gamma}$ could be probed by photon fusion at Belle II by searching 
for $e^+ e^- \to e^+ e^- + \phi\phi$ \cite{Dolan:2017osp}. The sensitivity for the axion-like particle coupling to photons, $ (g_{a\gamma\gamma}/4) a F_{\mu\nu} \tilde F^{\mu\nu}$, is estimated to be around the order of $g_{a\gamma\gamma}\sim 10^{-4}\,\rm GeV^{-1}$ for $m_a >150\,\rm MeV$ \cite{Dolan:2017osp}.
A rough translation of the bound in our case can be done by endowing one of the two $\phi$ states with a constant value $\phi_0$.
To be conservative, we take $\phi_0=10\,\rm GeV$ corresponding to the Belle II collision energy. 
Then $\phi_0 C_{\gamma\phi}\sim g_{a\gamma\gamma}/4$ results in  $m_Q \sim 11\,\rm GeV$ for ${\cal O}(1)$ Yukawa couplings, a comparable sensitivity as we obtained above.

\section{Discussion}
\label{sec:discussion}

The decays $B\to X_s\gamma$, $B_s\to \phi\phi$, $D^0\to \pi^0 + \phi\phi$, $D^0\to \phi\phi$ and $B_s-\bar B_s$ mixing can also be generated in the model and thus could pose constraints on the allowed 
parameter space that solves the Belle II anomaly. In the following, we estimate their effect and also comment on direct searches for vector-like quarks. 

\pmb{$B\to X_s\gamma$}: 
This transition is related to the dipole operators in the LEFT, 
$O_{d\gamma}^{ij} = \bar d_i \sigma^{\mu\nu}P_R d_j F_{\mu\nu}$, 
with the WC from the one-loop calculation given by
\begin{align}
\label{eq:Cdgamma}
C_{d\gamma}^{ij} = 
\frac{1}{16\pi^2 } \frac{e}{24} 
\left[
\frac{v y_q^i y_d^j y_1}{\sqrt{2} m_D m_Q } I(t)
+ \frac{5 m_{d_j} y_q^i y_q^{j*}}{6 m_Q^2}
+ \frac{ m_{d_i} y_d^{i*} y_d^j}{3 m_D^2}
\right],
\quad
I(t)  = \frac{5 - 7 t + 2 t^2 + 3 t \ln t}{(1- t)^2},
\end{align}
with $t\equiv m_Q^2/m_D^2$. 
Denoting the  term in square brackets in \cref{eq:Cdgamma} as $\tilde C_{d\gamma}^{ij}$, recent results of global fits 
\cite{Bause:2022rrs} result in the constraint $\tilde C_{d\gamma}^{sb} \lesssim 260/(10^5\,{\rm TeV})$. 
The quantity $\tilde C_{d\gamma}^{sb}/I(t)$ is roughly the same order as $ C_{d\phi}^{S,sb}$ given in \cref{eq:leftWCmatching}, and since 
the loop function $I(t)$ ranges between  5 and 2 as $t$ varies between zero and infinity, 
this implies that  $B\to X_s\gamma$ does not restrict our explanation of $B^+\to K^+ + \mathrm{inv}$, which requires the much smaller value $ C_{d\phi}^{S,sb}\sim (3-8)/(10^5\,{\rm TeV})$. Similarly, the related operator with a gluon field strength tensor is produced in our model with a coefficient,
\begin{align}
{\cal O}_{dG}^{ij}= \bar d_i T^A \sigma^{\mu\nu}P_R d_j G^A_{\mu\nu}, \quad 
C_{dG}^{ij}  = - {1 \over 16\pi^2}{g_s\over 4}
 \left( {y_q^i y_d^j y_1 v \over \sqrt{2} m_D m_Q }
 + {m_{d_j} y_q^j y_q^{j*} \over 6m_Q^2}
 + { m_{d_i} y_d^{i*} y_d^j \over 6m_D^2} \right).
\end{align}
The same global fits \cite{Bause:2022rrs} imply a constraint on the term in brackets about ten times weaker than its counterpart in \cref{eq:Cdgamma}.

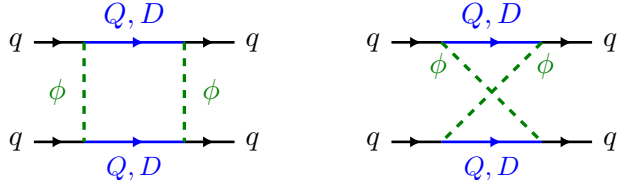
\begin{figure}
\centering
\begin{tikzpicture}[mystyle,scale=1.1]
\begin{scope}[shift={(1,1)}]
\draw[f] (-2,1) node[left]{$q$} -- (-1,1);
\draw[f,blue] (-1,1) -- (1,1) node[midway, yshift =  10 pt]{$Q,D$};
\draw[f] (1,1) -- (2,1)node[right]{$q$};
\draw[f] (-2,-1) node[left]{$q$} -- (-1,-1);
\draw[f,blue] (-1,-1) -- (1,-1) node[midway, yshift = - 10 pt]{\small $Q,D$};
\draw[f] (1,-1) -- (2,-1)node[right]{$q$};
\draw[snar, darkgreen] (-1,1) -- (-1,-1) node[midway, xshift = - 10 pt]{$\phi$};
\draw[snar, darkgreen] (1,1) -- (1,-1) node[midway, xshift = 10 pt]{$\phi$};
\end{scope}
\end{tikzpicture}\quad\quad\quad
\begin{tikzpicture}[mystyle,scale=1.1]
\begin{scope}[shift={(1,1)}]
\draw[f] (-2,1) node[left]{$q$} -- (-1,1);
\draw[f,blue] (-1,1) -- (1,1) node[midway, yshift =  10 pt]{$Q,D$};
\draw[f] (1,1) -- (2,1)node[right]{$q$};
\draw[f] (-2,-1) node[left]{$q$} -- (-1,-1);
\draw[f,blue] (-1,-1) -- (1,-1) node[midway, yshift = - 10 pt]{\small $Q,D$};
\draw[f] (1,-1) -- (2,-1)node[right]{$q$};
\draw[snar, darkgreen] (-1,1) -- (1,-1) node[midway, xshift = - 20 pt, yshift = 10 pt]{$\phi$};
\draw[snar, darkgreen] (1,1) -- (-1,-1) node[midway, xshift = 20 pt, yshift = 10 pt]{$\phi$};
\end{scope}
\end{tikzpicture}
\caption{Potential leading order Feynman diagrams contributing to neutral meson mixing.}
\label{fig:Feyn-dia-4q}
\end{figure}

\pmb{$B_s(s\bar b)-\bar B_s(\bar s b)$  mixing}:
Naively, we would expect the generation of dim-6 SMEFT four-quark operators mediating neutral meson mixing by the Feynman diagrams\footnote{Note that the penguin-like diagrams with a SM gauge boson or Higgs cannot yield such double flavour-changing operators.} shown in \cref{fig:Feyn-dia-4q}. 
However, due to the chirality of the quark pairs in each fermion line being the same, the sum of contributions from the two diagrams vanishes for dim-6 operators. Thus, the leading order SMEFT operators with four quarks which contribute to neutral meson mixing only arise at dim-8 order, with the structure $(\bar q q)^2\otimes\{D^2,G^A, H D, H^2\}$. Their WCs are the order $y_{\tt Yua}^6/(32 \pi^2 m_{Q,D}^4)$ with $y_{\tt Yua}$ representing the Yukawa couplings in the model.
The largest contribution to $B_s-\bar B_s$ mixing comes from the operator with field content $(\bar q q)^2H^2$ , while other contributions are suppressed at low energy by the ratio of the hadronic scale to the Higgs vev. The dominant 
operator can be generated by diagrams such as those in \cref{fig:Feyn-dia-4q} but with additional Higgs fields attached to the internal heavy vector quark lines, and results  in 
$y_{\tt Yua}^6\sim y_q^2 y_d^2 y_{1,2}^2$.
After electroweak symmetry breaking, the WCs for the four-quark operators $(\bar s \Gamma_1 b)(\bar s\Gamma_2 b)$ are estimated to be 
\begin{align}
C_{sbsb} & \sim \frac{[(y_q^s y_d^b)^2, (y_d^{s*} y_q^{b*})^2, y_q^s y_d^{s*} y_q^{b*} y_d^b] y_{1,2}^2 }{32\pi^2 m_{Q,D}^4} \frac{v^2}{8}
\nonumber
\\
& \approx 
3 \times 10^{-8} G_F \left(  3\,{\rm TeV} \over m_{Q,D} \right)^4 
[(y_q^s y_d^b)^2, (y_d^{s*} y_q^{b*})^2, y_q^s y_d^{s*} y_q^{b*} y_d^b] y_{1,2}^2, 
\end{align}
where the three different Yukawa combinations correspond to the three different operator structures: $(\bar s (1-\gamma_5) b)(\bar s(1-\gamma_5) b)$, 
$(\bar s (1+\gamma_5) b)(\bar s(1+\gamma_5) b)$, 
and 
$(\bar s (1-\gamma_5) b)(\bar s(1+\gamma_5) b)$, respectively.  
For the SM case, the WC for the operator $(\bar s\gamma_\mu(1-\gamma_5)b)(\bar s\gamma^\mu(1-\gamma_5)b)$
is $1.9\times 10^{-6} G_F$. We can see the new physics contribution is strongly suppressed relative to the SM once the NP scale is higher than 3 TeV for an ${\cal O}(1)$ coupling constant.
From \cref{fig:btokss} and \cref{eq:leftWCmatching} 
the $B^+ \to K^+ + \mathrm{inv}$ excess requires a new physics scale around $50\, \rm TeV$ for ${\cal O}(1)$ coupling constant,
and thus the constraints from $B_s-\bar B_s$ mixing can be neglected. 
For $B-\bar B$ and $K-\bar K$ mixing, the new contributions take a similar form as above with a change of corresponding flavor labels, $y_{d,q}^s\to y_{d,q}^d$ and $y_{d,q}^b\to y_{d,q}^d$, respectively.  
From the choice $|y_{q,d}^d| \sim 0.2 |y_{d,s}^s|$, it follows that there is available parameter space for other parameters to satisfy the constraints from these processes. For instance, by taking $y_1=1, m_{D,Q}={3\,\rm TeV}, y_{q,d}^s = 2$ and $y_2 \approx -2.582 y_1 $, we find the mass difference for kaon mixing is $\Delta M\sim 6.2\times 10^{-16}\, \rm GeV$, which is smaller by an order of magnitude than the experimental uncertainty.

\pmb{$B_s\to \mathrm{inv}$}:
Its branching ratio takes the form
\begin{align}
{\cal B}(B_s \to \phi\phi) = 
\frac{ |C_{d\phi}^{P,sb}|^2 \tau_{B_s} m_{B_s} f_{B_s}^2}{32\pi }    
\left( {m_{B_s} \over m_b +m_s} \right)^2
\sqrt{1 - 4 m_\phi^2 / m_{B_s}^2}.
\end{align}
Similar to $B\to K^*\phi\phi$ it depends on the pseudo-scalar WC $C^{P,sb}_{d\phi}$ and thus the current constraint of $\mathcal{B}(B_s\to \mathrm{inv}) < 5.4\times 10^{-4}$ at 90\% c.l.~\cite{Alonso-Alvarez:2023mgc} is easily satisfied with our choice of the Yukawa couplings. For $m_\phi=0.6$ GeV and $f_{B_s}=225.3(6.6)$ MeV \cite{FlavourLatticeAveragingGroupFLAG:2021npn},
we find a branching ratio 
${\cal B}(B_s \to \phi\phi) \sim 4\times 10^{-5}$
for $|C_{d\phi}^{S,ss}|\sim 0.13/\rm TeV$ and $\sqrt{r_d^2+r_q^2 - 2r_d r_q\cos\theta}\sim 10^{-3}$ in \cref{eq:Cdphimodel} that can simultaneously satisfy   
the DM relic density and $B^+\to K^++\mathrm{inv}$ requirements. At this level, the rate is about four times larger than the expected sensitivity at Belle II with 5~ab$^{-1}$ \cite{Belle-II:2018jsg}.

\pmb{$D^0 \to \mathrm{inv}$ and $D^0 \to \pi^0 +\mathrm{inv}$ decays}: 
These decay modes have been measured by Belle and BESIII respectively yielding 90\% c.l.\,\,upper limits 
${\cal B}(D^0 \to \mathrm{inv}) < 9.4\times 10^{-5}$ \cite{Belle:2016qek} 
and ${\cal B}(D^0 \to\pi^0 \nu\bar\nu) < 2.1\times 10^{-4}$ \cite{BESIII:2021slf}. 
These decays are induced by the operators ${\cal O}_{u\phi}^{S(P),uc}$ with up and charm quark flavours and could therefore constrain the corresponding WCs. Reference~\cite{Li:2023sjf} has recently studied the constraints on 
${\cal O}^{S(P),uc}_{u\phi}$ from these modes (see Eq.~2 and Fig.~5 in that paper). In our notation, those constraints translate into $|C_{u\phi}^{S(P),uc}|/m_c \lesssim 2.06(0.25)/\rm TeV^2$ for $m_\phi = 0.5\,\rm GeV$ and become weaker for larger $m_\phi$. These bounds are satisfied by the benchmark parameters used below, for which   
$|C_{u\phi}^{S(P),uc}|/m_c 
\simeq (0.05/\,{\rm TeV}^2)\left( {|y_q^s| \over 2} { 3\,{\rm TeV} \over m_Q} \right)^2$  
after neglecting the Yukawa couplings associated with first and third-generation quarks. 

\pmb{$h \to gg$, $h\to \gamma\gamma$, $Z\to b\bar b$, and $Z\to \mathrm{inv}$ decays}:
There are additional contributions to $h \to gg$ and  $h\to \gamma\gamma$ that can be
induced by double insertion of the Yukawa terms with $y_{1,2}$ in \cref{eq:modelL} at one-loop level. For the gluon case, both the CP even and CP odd operators can be generated, 
\begin{align}
\Delta {\cal L}_{h\to gg} = - { {\rm Re}[y_1y_2^*]v^2 \over m_D m_Q} \left[  {\alpha_s\over 12\pi} {h \over v}G_{\mu\nu}^A G^{A\mu\nu} \right]- {\alpha_s\over 4\pi} { {\rm Im}[y_1y_2^*]v \over 2 m_D m_Q}  hG_{\mu\nu}^A \tilde G^{A\mu\nu}, 
\end{align}
where the square bracket term represents the one-loop SM contribution from a top quark loop. 
It can be seen the correction is small compared to the top quark contribution, $y_1 y_2^* v^2/(m_Q m_D)\sim 0.007y_2^*$ with $y_1=1$ and $m_D=m_Q=3\,\rm TeV$ as chosen below, and thus does not pose any relevant constraint. Moreover, the explanation of the dark matter abundance and $B^+\to K^+ +\mathrm{inv}$ is independent of the Yukawa coupling $y_2$ and thus the Higgs digluon coupling is not able to constrain the scenario. The contribution to Higgs diphoton decay will be suppressed similarly to the Higgs digluon coupling. 
Similarly, $Z\to b\bar b$ does not pose any meaningful constraint given our suppressed coupling to bottom quarks.
The decay $Z\to \mathrm{inv}$ through $Z\to \phi\phi$ is absent in our case due to Bose statistics for the identical 
final DM states and angular momentum conservation. 

\pmb{Direct searches at the LHC}: 
Existing searches~\cite{ATLAS:2022tla,CMS:2022fck} for pair-produced vector-like quarks at the LHC rely on their decays into top (or bottom) quarks associated with a $Z,W$ or Higgs boson. The  $\mathbb{Z}_2$ symmetry imposed on our model implies that their decays into quarks are accompanied instead by the light dark scalar. 
Arina et al.~\cite{Arina:2023msd} recently recast the CMS analysis within supersymmetry~\cite{CMS:2019zmd} to obtain bounds for dark matter models with a $t$-channel mediator~\cite{Arina:2020udz}. For real scalar dark matter with a fermionic mediator coupling to right-handed up quarks, the study obtained a lower bound of $1.5$ TeV on the mediator mass if the pair production is dominated by QCD which is consistent with the searches for vector-like quarks~\cite{ATLAS:2022tla,CMS:2022fck}. For large portal couplings, the production is dominated by $t$-channel scattering. For a portal coupling $\sim 5$, 
Ref.~\cite{Arina:2023msd} obtained a lower limit of $3.3$ TeV on the mediator mass. As such, the existing bounds do not apply and we leave an extraction of precise LHC bounds on these particles to a future analysis.
\begin{figure}
\centering
\includegraphics[width=0.7\linewidth]{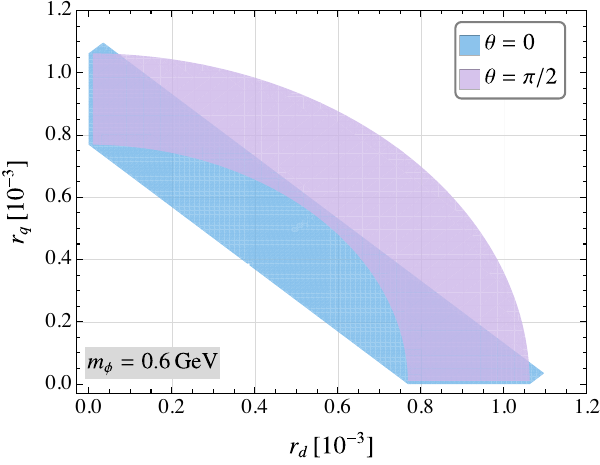}
\caption{The allowed parameter space for DM mass $m_\phi =0.6\,\rm GeV$, $m_Q=m_D=3\,\rm TeV$ and two benchmark CP phases as function of the coupling ratios $r_{d,q}=|y_{d,q}^b|/|y_{d,q}^s|$.}
\label{fig:model}
\end{figure}
\pmb{The viable parameter space}:
We are now in a position to discuss the viable parameter space which explains the excess in $B^+ \to K^+ + \mathrm{inv}$ and provides the correct DM relic density, while satisfying various constraints, in particular those posed by $B \to K^* + \mathrm{inv}$.
Firstly, due to the suppression of the second terms of $C_{d\phi}^{S(P),ij}$ in \cref{eq:leftWCmatching} by SM quark masses, we neglect their contribution and focus on 
the analysis of the first terms. 
By decomposing the complex Yukawa couplings as 
$y_{q}^i \equiv |y_{q}^i| e^{- i \alpha_i}$ and 
$y_{d}^i \equiv |y_{d}^i| e^{- i \beta_i}$, 
with $\alpha_i$ and $\beta_i$ being the associated CP phases,
the absolute values of the three relevant WCs become, 
\begin{subequations}
\begin{align}
|C_{d\phi}^{S,ss}| & \approx
\frac{ |y_{q}^s| |y_{d}^s| y_1 v}{\sqrt{2}m_Q m_D }
\left|  1 + e^{i 2(\alpha_s + \beta_s )} \right|, 
\\
|C_{d\phi}^{S(P),sb}| & \approx
\frac{|y_{q}^s| |y_{d}^s| y_1 v}{\sqrt{2}m_Q m_D }
\left|  { |y_{d}^b | \over |y_{d}^s| }  \pm { |y_{q}^b| \over |y_{q}^s| }  e^{i (\alpha_s + \alpha_b + \beta_s+ \beta_b )}
\right|.
\end{align}
\end{subequations}
Next we assume the CP phases $\alpha_s+ \beta_s =0$
and denote
$\alpha_b+\beta_b \equiv \theta$, 
$|y_{d}^b |/|y_{d}^s| \equiv r_d$,
$|y_{q}^b |/|y_{q}^s| \equiv r_q$. 
Then the above WCs obey the approximate relation, 
\begin{align}
\label{eq:Cdphimodel}
 |C_{d\phi}^{S(P),sb}| \approx {1\over 2}|C_{d\phi}^{S,ss}| \sqrt{r_d^2 +r_q^2 \pm 2r_d r_q \cos \theta}. 
\end{align}

From \cref{fig:dmabundance} we see that for the DM mass in the range $(0.5-0.9)\,\rm GeV$, the required WC $|C_{d\phi}^{S,ss}|_{\tt DM}\sim 0.12/\rm TeV$ is roughly independent of that mass. 
To achieve this value, we choose the masses for the two heavy quarks to be $m_Q = m_D = 3\,\rm TeV$, the Yukawa couplings related to the SM quarks to be $|y_{d}^s|=|y_{q}^s|=2$, and the one inducing heavy quark mixing to be $y_1=1$, leading to a comparable value for DM relic requirement, $|C_{d\phi}^{S,ss}|\sim 0.16/\rm TeV$. The light-blue (light-purple) region in \cref{fig:model} shows the allowed parameter space in $r_d-r_q$ plane for the CP phase $\theta=0$ ($\theta=\pi/2$) with the DM mass $m_\phi = 0.6\,\rm GeV$.\footnote{One of the observable effects of the CP violating phase is the future precision measurement of neutral meson mixing such as $K-\bar K$ and $B_{d,s}-\bar B_{d,s}$. }
It can be seen that the allowed regions require an ${\cal O}(10^{-3})$ hierarchy between the 
third and second generation Yukawa couplings, as observed before. 
This hierarchical structure could be flattened
if the DM is not mainly produced through $K$ and $\eta$ annihilation channels but includes additional channels like annihilation to pions through couplings to the down quark or annihilation via the Higgs portal interaction. 

\section{Conclusions}
\label{sec:conclusions}

In this paper, we introduce a UV complete model with two vector-like quarks $Q\sim(\mathbf{3},\mathbf{2},1/6)$ and $D\sim(\mathbf{3},\mathbf{1},-1/3)$ together with a real scalar singlet $\phi$ which are all odd with respect to a $\mathbb{Z}_2$ symmetry. Together they can enhance the branching ratio of $B \to K+\text{invisible}$ through additional decays to dark matter pairs to explain the recent Belle II excess and simultaneously generate the observed dark matter relic density through the thermal freeze-out mechanism via DM pair annihilation into meson pairs, $\phi\phi \longleftrightarrow KK(\eta\eta)$.   
Meanwhile, the model satisfies the strong constraints from $B_s - \bar{B}_s$ mixing, meson decays and dark matter (in-)direct detection. In particular, the related constraints from the associated $B \to K^* + \text{invisible}$ decays and invisible $B_s$ decays may be easily satisfied by choosing the Yukawa couplings so that the scalar operator dominates over the pseudo-scalar
operator, which is achieved for similar couplings to left-handed and right-handed down-type quarks, $y_d^i\simeq y_q^{i*}$. 

We are not aware of any dedicated searches for vector-like quarks which decay to quarks and a DM particle at the LHC, but anticipate lower bounds of order of $\mathcal{O}(1.5\ \mathrm{TeV})$ based on existing searches for vector-like quarks that decay with missing energy~\cite{ATLAS:2022tla,CMS:2022fck}. Note, however, that this bound was obtained with the requirement of leptons and additional jets in the final state, so is not directly applicable to our model. A dedicated search will provide important additional information and may be able to confirm or rule out the explanation of the dark matter abundance.

\section*{Acknowledgements}

This work was supported in part by Australian Research Council Discovery Project DP200101470 and in part by the Australian Research Council Centre of Excellence for Dark Matter Particle Physics (CDM, CE200100008). 
This work was also supported in part by Key Laboratory for Particle Physics, Astrophysics and Cosmology, Ministry of Education, and Shanghai Key Laboratory for Particle Physics and Cosmology (Grant No. 15DZ2272100) and in part by the NSFC (Grant Nos. 12375088, 12090064, and 12090060), and by the Guangdong Major Project of Basic and Applied Basic Research ( Grant No.2020B0301030008) and by the NSFC (Grant No. NSFC-12305110). XDM would like to thank Jin-Han Liang and Hao-Lin Wang 
for useful discussions about the Migdal effect. RRV thanks Peter Cox for a helpful discussion. 

\setlength{\bibsep}{.2\baselineskip plus 0.3ex}
\bibliographystyle{utphys} 
\bibliography{refs}

\providecommand{\href}[2]{#2}\begingroup\raggedright\begin{thebibliography}{10}

\bibitem{Belle-II:2023esi}
{\bfseries Belle-II} Collaboration, I.~Adachi {\em et~al.}, ``{Evidence for
  $B^{+}\to K^{+}\nu\bar{\nu}$ Decays},''
  \href{https://arxiv.org/abs/2311.14647}{{\ttfamily arXiv:2311.14647
  [hep-ex]}}.

\bibitem{Becirevic:2023aov}
D.~Be\v{c}irevi\'c, G.~Piazza, and O.~Sumensari, ``{Revisiting $B\rightarrow
  K^{(*)} \nu {\bar{\nu }}$ decays in the Standard Model and beyond},''
  \href{https://dx.doi.org/10.1140/epjc/s10052-023-11388-z}{{\em Eur. Phys. J.
  C} {\bfseries 83} no.~3, (2023) 252},
  \href{https://arxiv.org/abs/2301.06990}{{\ttfamily arXiv:2301.06990
  [hep-ph]}}.

\bibitem{Belle-II:2021rof}
{\bfseries Belle-II} Collaboration, F.~Abudin\'en {\em et~al.}, ``{Search for
  $B^+\to K^+ \nu\bar\nu$ Decays Using an Inclusive Tagging Method at Belle
  II},'' \href{https://dx.doi.org/10.1103/PhysRevLett.127.181802}{{\em Phys.
  Rev. Lett.} {\bfseries 127} no.~18, (2021) 181802},
  \href{https://arxiv.org/abs/2104.12624}{{\ttfamily arXiv:2104.12624
  [hep-ex]}}.

\bibitem{Belle:2013tnz}
{\bfseries Belle} Collaboration, O.~Lutz {\em et~al.}, ``{Search for $B \to
  h^{(*)} \nu \bar{\nu}$ with the full Belle $\Upsilon(4S)$ data sample},''
  \href{https://dx.doi.org/10.1103/PhysRevD.87.111103}{{\em Phys. Rev. D}
  {\bfseries 87} no.~11, (2013) 111103},
  \href{https://arxiv.org/abs/1303.3719}{{\ttfamily arXiv:1303.3719 [hep-ex]}}.

\bibitem{Belle:2017oht}
{\bfseries Belle} Collaboration, J.~Grygier {\em et~al.}, ``{Search for
  $\boldsymbol{B\to h\nu\bar{\nu}}$ decays with semileptonic tagging at
  Belle},'' \href{https://dx.doi.org/10.1103/PhysRevD.96.091101}{{\em Phys.
  Rev. D} {\bfseries 96} no.~9, (2017) 091101},
  \href{https://arxiv.org/abs/1702.03224}{{\ttfamily arXiv:1702.03224
  [hep-ex]}}. [Addendum: Phys.Rev.D 97, 099902 (2018)].

\bibitem{BaBar:2010oqg}
{\bfseries BaBar} Collaboration, P.~del Amo~Sanchez {\em et~al.}, ``{Search for
  the Rare Decay $B \to K \nu \bar{\nu}$},''
  \href{https://dx.doi.org/10.1103/PhysRevD.82.112002}{{\em Phys. Rev. D}
  {\bfseries 82} (2010) 112002},
  \href{https://arxiv.org/abs/1009.1529}{{\ttfamily arXiv:1009.1529 [hep-ex]}}.

\bibitem{BaBar:2013npw}
{\bfseries BaBar} Collaboration, J.~P. Lees {\em et~al.}, ``{Search for $B \to
  K^{(*)} \nu \overline \nu$ and invisible quarkonium decays},''
  \href{https://dx.doi.org/10.1103/PhysRevD.87.112005}{{\em Phys. Rev. D}
  {\bfseries 87} no.~11, (2013) 112005},
  \href{https://arxiv.org/abs/1303.7465}{{\ttfamily arXiv:1303.7465 [hep-ex]}}.

\bibitem{Bause:2023mfe}
R.~Bause, H.~Gisbert, and G.~Hiller, ``{Implications of an enhanced $B\to
  K\nu\bar\nu$ branching ratio},''
  \href{https://dx.doi.org/10.1103/PhysRevD.109.015006}{{\em Phys. Rev. D}
  {\bfseries 109} no.~1, (2024) 015006},
  \href{https://arxiv.org/abs/2309.00075}{{\ttfamily arXiv:2309.00075
  [hep-ph]}}.

\bibitem{Athron:2023hmz}
P.~Athron, R.~Martinez, and C.~Sierra, ``{B meson anomalies and large $
  {B}^{+}\to {K}^{+}\nu \overline{\nu} $ in non-universal $U(1)^\prime$
  models},'' \href{https://dx.doi.org/10.1007/JHEP02(2024)121}{{\em JHEP}
  {\bfseries 02} (2024) 121},
  \href{https://arxiv.org/abs/2308.13426}{{\ttfamily arXiv:2308.13426
  [hep-ph]}}.

\bibitem{Allwicher:2023xba}
L.~Allwicher, D.~Becirevic, G.~Piazza, S.~Rosauro-Alcaraz, and O.~Sumensari,
  ``{Understanding the first measurement of B($B\to K\nu\bar\nu$)},''
  \href{https://dx.doi.org/10.1016/j.physletb.2023.138411}{{\em Phys. Lett. B}
  {\bfseries 848} (2024) 138411},
  \href{https://arxiv.org/abs/2309.02246}{{\ttfamily arXiv:2309.02246
  [hep-ph]}}.

\bibitem{Amhis:2023mpj}
Y.~Amhis, M.~Kenzie, M.~Reboud, and A.~R. Wiederhold, ``{Prospects for searches
  of $ b\to s\nu \overline{\nu} $ decays at FCC-ee},''
  \href{https://dx.doi.org/10.1007/JHEP01(2024)144}{{\em JHEP} {\bfseries 01}
  (2024) 144}, \href{https://arxiv.org/abs/2309.11353}{{\ttfamily
  arXiv:2309.11353 [hep-ex]}}.

\bibitem{He:2023bnk}
X.-G. He, X.-D. Ma, and G.~Valencia, ``{Revisiting models that enhance $B^+\to
  K^+ \nu\bar\nu$ in light of the new Belle II measurement},''
  \href{https://arxiv.org/abs/2309.12741}{{\ttfamily arXiv:2309.12741
  [hep-ph]}}.

\bibitem{Chen:2023wpb}
C.-H. Chen and C.-W. Chiang, ``{Flavor anomalies in leptoquark model with
  gauged $U(1)_{L_\mu-L_\tau}$},''
  \href{https://arxiv.org/abs/2309.12904}{{\ttfamily arXiv:2309.12904
  [hep-ph]}}.

\bibitem{Fridell:2023ssf}
K.~Fridell, M.~Ghosh, T.~Okui, and K.~Tobioka, ``{Decoding the $B \to K \nu
  \nu$ excess at Belle II: kinematics, operators, and masses},''
  \href{https://arxiv.org/abs/2312.12507}{{\ttfamily arXiv:2312.12507
  [hep-ph]}}.

\bibitem{Chen:2024jlj}
F.-Z. Chen, Q.~Wen, and F.~Xu, ``{Correlating $B\to K^{(*)} \nu\bar{\nu}$ and
  flavor anomalies in SMEFT},''
  \href{https://arxiv.org/abs/2401.11552}{{\ttfamily arXiv:2401.11552
  [hep-ph]}}.

\bibitem{Loparco:2024olo}
F.~Loparco, ``{A new look at $b \to s$ observables in 331 models},''
  \href{https://arxiv.org/abs/2401.11999}{{\ttfamily arXiv:2401.11999
  [hep-ph]}}.

\bibitem{Hou:2024vyw}
B.-F. Hou, X.-Q. Li, M.~Shen, Y.-D. Yang, and X.-B. Yuan, ``{Deciphering the
  Belle II data on $B\to K \nu \bar\nu$ decay in the (dark) SMEFT with minimal
  flavour violation},'' \href{https://arxiv.org/abs/2402.19208}{{\ttfamily
  arXiv:2402.19208 [hep-ph]}}.

\bibitem{Chen:2024cll}
C.-H. Chen and C.-W. Chiang, ``{Rare $B$ and $K$ decays in a scotogenic
  model},'' \href{https://arxiv.org/abs/2403.02897}{{\ttfamily arXiv:2403.02897
  [hep-ph]}}.

\bibitem{Berezhnoy:2023rxx}
A.~Berezhnoy and D.~Melikhov, ``{$B\to K^* M_X$ vs $B\to K M_X$ as a probe of a
  scalar-mediator dark matter scenario},''
  \href{https://dx.doi.org/10.1209/0295-5075/ad1d03}{{\em EPL} {\bfseries 145}
  no.~1, (2024) 14001}, \href{https://arxiv.org/abs/2309.17191}{{\ttfamily
  arXiv:2309.17191 [hep-ph]}}.

\bibitem{Abdughani:2023dlr}
M.~Abdughani and Y.~Reyimuaji, ``{Constraining light dark matter and mediator
  with $B^+ \rightarrow K^+ \nu \bar \nu$ data},''
  \href{https://arxiv.org/abs/2309.03706}{{\ttfamily arXiv:2309.03706
  [hep-ph]}}.

\bibitem{Datta:2023iln}
A.~Datta, D.~Marfatia, and L.~Mukherjee, ``{$B\to K\nu\bar\nu$, MiniBooNE and
  muon g-2 anomalies from a dark sector},''
  \href{https://dx.doi.org/10.1103/PhysRevD.109.L031701}{{\em Phys. Rev. D}
  {\bfseries 109} no.~3, (2024) L031701},
  \href{https://arxiv.org/abs/2310.15136}{{\ttfamily arXiv:2310.15136
  [hep-ph]}}.

\bibitem{Altmannshofer:2023hkn}
W.~Altmannshofer, A.~Crivellin, H.~Haigh, G.~Inguglia, and J.~Martin~Camalich,
  ``{Light New Physics in $B\to K^{(*)}\nu\bar\nu$?},''
  \href{https://arxiv.org/abs/2311.14629}{{\ttfamily arXiv:2311.14629
  [hep-ph]}}.

\bibitem{McKeen:2023uzo}
D.~McKeen, J.~N. Ng, and D.~Tuckler, ``{Higgs Portal Interpretation of the
  Belle II $B^+ \to K^+ \nu \nu$ Measurement},''
  \href{https://arxiv.org/abs/2312.00982}{{\ttfamily arXiv:2312.00982
  [hep-ph]}}.

\bibitem{Ho:2024cwk}
S.-Y. Ho, J.~Kim, and P.~Ko, ``{Recent $B^+ \!\to K^+\nu\bar{\nu}$ Excess and
  Muon $g-2$ Illuminating Light Dark Sector with Higgs Portal},''
  \href{https://arxiv.org/abs/2401.10112}{{\ttfamily arXiv:2401.10112
  [hep-ph]}}.

\bibitem{Gabrielli:2024wys}
E.~Gabrielli, L.~Marzola, K.~M\"u\"ursepp, and M.~Raidal, ``{Explaining the
  $B^+\to K^+ \nu \bar{\nu}$ excess via a massless dark photon},''
  \href{https://arxiv.org/abs/2402.05901}{{\ttfamily arXiv:2402.05901
  [hep-ph]}}.

\bibitem{Felkl:2023ayn}
T.~Felkl, A.~Giri, R.~Mohanta, and M.~A. Schmidt, ``{When energy goes missing:
  new physics in $b\rightarrow s \nu \nu $ with sterile neutrinos},''
  \href{https://dx.doi.org/10.1140/epjc/s10052-023-12326-9}{{\em Eur. Phys. J.
  C} {\bfseries 83} no.~12, (2023) 1135},
  \href{https://arxiv.org/abs/2309.02940}{{\ttfamily arXiv:2309.02940
  [hep-ph]}}.

\bibitem{Dreiner:2023cms}
H.~K. Dreiner, J.~Y. G\"unther, and Z.~S. Wang, ``{The Decay $B\to
  K\nu\bar{\nu}$ at Belle II and a Massless Bino in R-parity-violating
  Supersymmetry},'' \href{https://arxiv.org/abs/2309.03727}{{\ttfamily
  arXiv:2309.03727 [hep-ph]}}.

\bibitem{Browder:2021hbl}
T.~E. Browder, N.~G. Deshpande, R.~Mandal, and R.~Sinha, ``{Impact of
  B\textrightarrow{}K\ensuremath{\nu}\ensuremath{\nu}\textasciimacron{}
  measurements on beyond the Standard Model theories},''
  \href{https://dx.doi.org/10.1103/PhysRevD.104.053007}{{\em Phys. Rev. D}
  {\bfseries 104} no.~5, (2021) 053007},
  \href{https://arxiv.org/abs/2107.01080}{{\ttfamily arXiv:2107.01080
  [hep-ph]}}.

\bibitem{He:2021yoz}
X.~G. He and G.~Valencia, ``{RK(\textasteriskcentered{})\ensuremath{\nu} and
  non-standard neutrino interactions},''
  \href{https://dx.doi.org/10.1016/j.physletb.2021.136607}{{\em Phys. Lett. B}
  {\bfseries 821} (2021) 136607},
  \href{https://arxiv.org/abs/2108.05033}{{\ttfamily arXiv:2108.05033
  [hep-ph]}}.

\bibitem{Felkl:2021uxi}
T.~Felkl, S.~L. Li, and M.~A. Schmidt, ``{A tale of invisibility: constraints
  on new physics in b \textrightarrow{} s\ensuremath{\nu}\ensuremath{\nu}},''
  \href{https://dx.doi.org/10.1007/JHEP12(2021)118}{{\em JHEP} {\bfseries 12}
  (2021) 118}, \href{https://arxiv.org/abs/2111.04327}{{\ttfamily
  arXiv:2111.04327 [hep-ph]}}.

\bibitem{He:2022ljo}
X.-G. He, X.-D. Ma, and G.~Valencia, ``{FCNC B and K meson decays with light
  bosonic Dark Matter},''
  \href{https://dx.doi.org/10.1007/JHEP03(2023)037}{{\em JHEP} {\bfseries 03}
  (2023) 037}, \href{https://arxiv.org/abs/2209.05223}{{\ttfamily
  arXiv:2209.05223 [hep-ph]}}.

\bibitem{Ovchynnikov:2023von}
M.~Ovchynnikov, M.~A. Schmidt, and T.~Schwetz, ``{Complementarity of
  $B\rightarrow K^{(*)} \mu \bar{\mu }$ and $B\rightarrow K^{(*)} +
  \textrm{inv}$ for searches of GeV-scale Higgs-like scalars},''
  \href{https://dx.doi.org/10.1140/epjc/s10052-023-11975-0}{{\em Eur. Phys. J.
  C} {\bfseries 83} no.~9, (2023) 791},
  \href{https://arxiv.org/abs/2306.09508}{{\ttfamily arXiv:2306.09508
  [hep-ph]}}.

\bibitem{Asadi:2023ucx}
P.~Asadi, A.~Bhattacharya, K.~Fraser, S.~Homiller, and A.~Parikh, ``{Wrinkles
  in the Froggatt-Nielsen mechanism and flavorful new physics},''
  \href{https://dx.doi.org/10.1007/JHEP10(2023)069}{{\em JHEP} {\bfseries 10}
  (2023) 069}, \href{https://arxiv.org/abs/2308.01340}{{\ttfamily
  arXiv:2308.01340 [hep-ph]}}.

\bibitem{Bird:2004ts}
C.~Bird, P.~Jackson, R.~V. Kowalewski, and M.~Pospelov, ``{Search for dark
  matter in $b\to s$ transitions with missing energy},''
  \href{https://dx.doi.org/10.1103/PhysRevLett.93.201803}{{\em Phys. Rev.
  Lett.} {\bfseries 93} (2004) 201803},
  \href{https://arxiv.org/abs/hep-ph/0401195}{{\ttfamily
  arXiv:hep-ph/0401195}}.

\bibitem{ATLAS:2023tkt}
{\bfseries ATLAS} Collaboration, G.~Aad {\em et~al.}, ``{Combination of
  searches for invisible decays of the Higgs boson using 139 fb\ensuremath{-}1
  of proton-proton collision data at s=13 TeV collected with the ATLAS
  experiment},'' \href{https://dx.doi.org/10.1016/j.physletb.2023.137963}{{\em
  Phys. Lett. B} {\bfseries 842} (2023) 137963},
  \href{https://arxiv.org/abs/2301.10731}{{\ttfamily arXiv:2301.10731
  [hep-ex]}}.

\bibitem{ATLAS:2022yvh}
{\bfseries ATLAS} Collaboration, G.~Aad {\em et~al.}, ``{Search for invisible
  Higgs-boson decays in events with vector-boson fusion signatures using 139
  fb$^{-1}$ of proton-proton data recorded by the ATLAS experiment},''
  \href{https://dx.doi.org/10.1007/JHEP08(2022)104}{{\em JHEP} {\bfseries 08}
  (2022) 104}, \href{https://arxiv.org/abs/2202.07953}{{\ttfamily
  arXiv:2202.07953 [hep-ex]}}.

\bibitem{CMS:2022qva}
{\bfseries CMS} Collaboration, A.~Tumasyan {\em et~al.}, ``{Search for
  invisible decays of the Higgs boson produced via vector boson fusion in
  proton-proton collisions at s=13\,\,TeV},''
  \href{https://dx.doi.org/10.1103/PhysRevD.105.092007}{{\em Phys. Rev. D}
  {\bfseries 105} no.~9, (2022) 092007},
  \href{https://arxiv.org/abs/2201.11585}{{\ttfamily arXiv:2201.11585
  [hep-ex]}}.

\bibitem{Fuentes-Martin:2022jrf}
J.~Fuentes-Mart\'\i{}n, M.~K\"onig, J.~Pag\`es, A.~E. Thomsen, and F.~Wilsch,
  ``{A proof of concept for matchete: an automated tool for matching effective
  theories},'' \href{https://dx.doi.org/10.1140/epjc/s10052-023-11726-1}{{\em
  Eur. Phys. J. C} {\bfseries 83} no.~7, (2023) 662},
  \href{https://arxiv.org/abs/2212.04510}{{\ttfamily arXiv:2212.04510
  [hep-ph]}}.

\bibitem{Gubernari:2023puw}
N.~Gubernari, M.~Reboud, D.~van Dyk, and J.~Virto, ``{Dispersive analysis of B
  \textrightarrow{} K$^{(*)}$ and B$_{s}$\textrightarrow{} \ensuremath{\phi}
  form factors},'' \href{https://dx.doi.org/10.1007/JHEP12(2023)153}{{\em JHEP}
  {\bfseries 12} (2023) 153},
  \href{https://arxiv.org/abs/2305.06301}{{\ttfamily arXiv:2305.06301
  [hep-ph]}}.

\bibitem{Gartner:2024muk}
L.~G\"artner, N.~Hartmann, L.~Heinrich, M.~Horstmann, T.~Kuhr, M.~Reboud,
  S.~Stefkova, and D.~van Dyk, ``{Constructing model-agnostic likelihoods, a
  method for the reinterpretation of particle physics results},''
  \href{https://arxiv.org/abs/2402.08417}{{\ttfamily arXiv:2402.08417
  [hep-ph]}}.

\bibitem{Gasser:1983yg}
J.~Gasser and H.~Leutwyler, ``{Chiral Perturbation Theory to One Loop},''
  \href{https://dx.doi.org/10.1016/0003-4916(84)90242-2}{{\em Annals Phys.}
  {\bfseries 158} (1984) 142}.

\bibitem{Gasser:1984gg}
J.~Gasser and H.~Leutwyler, ``{Chiral Perturbation Theory: Expansions in the
  Mass of the Strange Quark},''
  \href{https://dx.doi.org/10.1016/0550-3213(85)90492-4}{{\em Nucl. Phys. B}
  {\bfseries 250} (1985) 465--516}.

\bibitem{Gondolo:1990dk}
P.~Gondolo and G.~Gelmini, ``{Cosmic abundances of stable particles: Improved
  analysis},'' \href{https://dx.doi.org/10.1016/0550-3213(91)90438-4}{{\em
  Nucl. Phys. B} {\bfseries 360} (1991) 145--179}.

\bibitem{Steigman:2012nb}
G.~Steigman, B.~Dasgupta, and J.~F. Beacom, ``{Precise Relic WIMP Abundance and
  its Impact on Searches for Dark Matter Annihilation},''
  \href{https://dx.doi.org/10.1103/PhysRevD.86.023506}{{\em Phys. Rev. D}
  {\bfseries 86} (2012) 023506},
  \href{https://arxiv.org/abs/1204.3622}{{\ttfamily arXiv:1204.3622 [hep-ph]}}.

\bibitem{Migdal:1941}
A.~B. Migdal, ``{Ionization of atoms accompanying $\alpha$-and
  $\beta$-decay},'' {\em J. Phys. 4, 449 (1941)} (1941) .

\bibitem{Ibe:2017yqa}
M.~Ibe, W.~Nakano, Y.~Shoji, and K.~Suzuki, ``{Migdal Effect in Dark Matter
  Direct Detection Experiments},''
  \href{https://dx.doi.org/10.1007/JHEP03(2018)194}{{\em JHEP} {\bfseries 03}
  (2018) 194}, \href{https://arxiv.org/abs/1707.07258}{{\ttfamily
  arXiv:1707.07258 [hep-ph]}}.

\bibitem{Dolan:2017xbu}
M.~J. Dolan, F.~Kahlhoefer, and C.~McCabe, ``{Directly detecting sub-GeV dark
  matter with electrons from nuclear scattering},''
  \href{https://dx.doi.org/10.1103/PhysRevLett.121.101801}{{\em Phys. Rev.
  Lett.} {\bfseries 121} no.~10, (2018) 101801},
  \href{https://arxiv.org/abs/1711.09906}{{\ttfamily arXiv:1711.09906
  [hep-ph]}}.

\bibitem{XENON:2019zpr}
{\bfseries XENON} Collaboration, E.~Aprile {\em et~al.}, ``{Search for Light
  Dark Matter Interactions Enhanced by the Migdal Effect or Bremsstrahlung in
  XENON1T},'' \href{https://dx.doi.org/10.1103/PhysRevLett.123.241803}{{\em
  Phys. Rev. Lett.} {\bfseries 123} no.~24, (2019) 241803},
  \href{https://arxiv.org/abs/1907.12771}{{\ttfamily arXiv:1907.12771
  [hep-ex]}}.

\bibitem{DarkSide:2022dhx}
{\bfseries DarkSide} Collaboration, P.~Agnes {\em et~al.}, ``{Search for
  Dark-Matter\textendash{}Nucleon Interactions via Migdal Effect with
  DarkSide-50},'' \href{https://dx.doi.org/10.1103/PhysRevLett.130.101001}{{\em
  Phys. Rev. Lett.} {\bfseries 130} no.~10, (2023) 101001},
  \href{https://arxiv.org/abs/2207.11967}{{\ttfamily arXiv:2207.11967
  [hep-ex]}}.

\bibitem{LZ:2023poo}
{\bfseries LZ} Collaboration, J.~Aalbers {\em et~al.}, ``{Search for new
  physics in low-energy electron recoils from the first LZ exposure},''
  \href{https://dx.doi.org/10.1103/PhysRevD.108.072006}{{\em Phys. Rev. D}
  {\bfseries 108} no.~7, (2023) 072006},
  \href{https://arxiv.org/abs/2307.15753}{{\ttfamily arXiv:2307.15753
  [hep-ex]}}.

\bibitem{PandaX:2023xgl}
{\bfseries PandaX} Collaboration, D.~Huang {\em et~al.}, ``{Search for
  Dark-Matter\textendash{}Nucleon Interactions with a Dark Mediator in
  PandaX-4T},'' \href{https://dx.doi.org/10.1103/PhysRevLett.131.191002}{{\em
  Phys. Rev. Lett.} {\bfseries 131} no.~19, (2023) 191002},
  \href{https://arxiv.org/abs/2308.01540}{{\ttfamily arXiv:2308.01540
  [hep-ex]}}.

\bibitem{DelNobile:2021wmp}
E.~Del~Nobile, ``{The Theory of Direct Dark Matter Detection: A Guide to
  Computations},'' \href{https://arxiv.org/abs/2104.12785}{{\ttfamily
  arXiv:2104.12785 [hep-ph]}}.

\bibitem{Liang:2024tef}
J.-H. Liang, Y.~Liao, X.-D. Ma, and H.-L. Wang, ``{Comprehensive constraints on
  fermionic dark matter-quark tensor interactions in direct detection
  experiments},'' \href{https://arxiv.org/abs/2401.05005}{{\ttfamily
  arXiv:2401.05005 [hep-ph]}}.

\bibitem{Boddy:2015efa}
K.~K. Boddy and J.~Kumar, ``{Indirect Detection of Dark Matter Using MeV-Range
  Gamma-Ray Telescopes},''
  \href{https://dx.doi.org/10.1103/PhysRevD.92.023533}{{\em Phys. Rev. D}
  {\bfseries 92} no.~2, (2015) 023533},
  \href{https://arxiv.org/abs/1504.04024}{{\ttfamily arXiv:1504.04024
  [astro-ph.CO]}}.

\bibitem{Dolan:2017osp}
M.~J. Dolan, T.~Ferber, C.~Hearty, F.~Kahlhoefer, and K.~Schmidt-Hoberg,
  ``{Revised constraints and Belle II sensitivity for visible and invisible
  axion-like particles},''
  \href{https://dx.doi.org/10.1007/JHEP12(2017)094}{{\em JHEP} {\bfseries 12}
  (2017) 094}, \href{https://arxiv.org/abs/1709.00009}{{\ttfamily
  arXiv:1709.00009 [hep-ph]}}. [Erratum: JHEP 03, 190 (2021)].

\bibitem{Bause:2022rrs}
R.~Bause, H.~Gisbert, M.~Golz, and G.~Hiller, ``{Model-independent analysis of
  $b \rightarrow d$ processes},''
  \href{https://dx.doi.org/10.1140/epjc/s10052-023-11586-9}{{\em Eur. Phys. J.
  C} {\bfseries 83} no.~5, (2023) 419},
  \href{https://arxiv.org/abs/2209.04457}{{\ttfamily arXiv:2209.04457
  [hep-ph]}}.

\bibitem{Alonso-Alvarez:2023mgc}
G.~Alonso-\'Alvarez and M.~Escudero, ``{The first limit on invisible decays of
  $B_s$ mesons comes from LEP},''
  \href{https://arxiv.org/abs/2310.13043}{{\ttfamily arXiv:2310.13043
  [hep-ph]}}.

\bibitem{FlavourLatticeAveragingGroupFLAG:2021npn}
{\bfseries Flavour Lattice Averaging Group (FLAG)} Collaboration, Y.~Aoki {\em
  et~al.}, ``{FLAG Review 2021},''
  \href{https://dx.doi.org/10.1140/epjc/s10052-022-10536-1}{{\em Eur. Phys. J.
  C} {\bfseries 82} no.~10, (2022) 869},
  \href{https://arxiv.org/abs/2111.09849}{{\ttfamily arXiv:2111.09849
  [hep-lat]}}.

\bibitem{Belle-II:2018jsg}
{\bfseries Belle-II} Collaboration, W.~Altmannshofer {\em et~al.}, ``{The Belle
  II Physics Book},'' \href{https://dx.doi.org/10.1093/ptep/ptz106}{{\em PTEP}
  {\bfseries 2019} no.~12, (2019) 123C01},
  \href{https://arxiv.org/abs/1808.10567}{{\ttfamily arXiv:1808.10567
  [hep-ex]}}. [Erratum: PTEP 2020, 029201 (2020)].

\bibitem{Belle:2016qek}
{\bfseries Belle} Collaboration, Y.~T. Lai {\em et~al.}, ``{Search for $D^{0}$
  decays to invisible final states at Belle},''
  \href{https://dx.doi.org/10.1103/PhysRevD.95.011102}{{\em Phys. Rev. D}
  {\bfseries 95} no.~1, (2017) 011102},
  \href{https://arxiv.org/abs/1611.09455}{{\ttfamily arXiv:1611.09455
  [hep-ex]}}.

\bibitem{BESIII:2021slf}
{\bfseries BESIII} Collaboration, M.~Ablikim {\em et~al.}, ``{Search for the
  decay $D^{0} \to \pi^{0} \nu \bar{\nu}$},''
  \href{https://dx.doi.org/10.1103/PhysRevD.105.L071102}{{\em Phys. Rev. D}
  {\bfseries 105} no.~7, (2022) L071102},
  \href{https://arxiv.org/abs/2112.14236}{{\ttfamily arXiv:2112.14236
  [hep-ex]}}.

\bibitem{Li:2023sjf}
G.~Li and J.~Tandean, ``{FCNC charmed-hadron decays with invisible singlet
  particles in light of recent data},''
  \href{https://dx.doi.org/10.1007/JHEP11(2023)205}{{\em JHEP} {\bfseries 11}
  (2023) 205}, \href{https://arxiv.org/abs/2306.05333}{{\ttfamily
  arXiv:2306.05333 [hep-ph]}}.

\bibitem{ATLAS:2022tla}
{\bfseries ATLAS} Collaboration, G.~Aad {\em et~al.}, ``{Search for
  pair-produced vector-like top and bottom partners in events with large
  missing transverse momentum in pp collisions with the ATLAS detector},''
  \href{https://dx.doi.org/10.1140/epjc/s10052-023-11790-7}{{\em Eur. Phys. J.
  C} {\bfseries 83} no.~8, (2023) 719},
  \href{https://arxiv.org/abs/2212.05263}{{\ttfamily arXiv:2212.05263
  [hep-ex]}}.

\bibitem{CMS:2022fck}
{\bfseries CMS} Collaboration, A.~Tumasyan {\em et~al.}, ``{Search for pair
  production of vector-like quarks in leptonic final states in proton-proton
  collisions at $ \sqrt{s} $ = 13 TeV},''
  \href{https://dx.doi.org/10.1007/JHEP07(2023)020}{{\em JHEP} {\bfseries 07}
  (2023) 020}, \href{https://arxiv.org/abs/2209.07327}{{\ttfamily
  arXiv:2209.07327 [hep-ex]}}.

\bibitem{Arina:2023msd}
C.~Arina, B.~Fuks, J.~Heisig, M.~Kr\"amer, L.~Mantani, and L.~Panizzi,
  ``{Comprehensive exploration of t-channel simplified models of dark
  matter},'' \href{https://dx.doi.org/10.1103/PhysRevD.108.115007}{{\em Phys.
  Rev. D} {\bfseries 108} no.~11, (2023) 115007},
  \href{https://arxiv.org/abs/2307.10367}{{\ttfamily arXiv:2307.10367
  [hep-ph]}}.

\bibitem{CMS:2019zmd}
{\bfseries CMS} Collaboration, T.~C. Collaboration {\em et~al.}, ``{Search for
  supersymmetry in proton-proton collisions at 13 TeV in final states with jets
  and missing transverse momentum},''
  \href{https://dx.doi.org/10.1007/JHEP10(2019)244}{{\em JHEP} {\bfseries 10}
  (2019) 244}, \href{https://arxiv.org/abs/1908.04722}{{\ttfamily
  arXiv:1908.04722 [hep-ex]}}.

\bibitem{Arina:2020udz}
C.~Arina, B.~Fuks, and L.~Mantani, ``{A universal framework for t-channel dark
  matter models},''
  \href{https://dx.doi.org/10.1140/epjc/s10052-020-7933-7}{{\em Eur. Phys. J.
  C} {\bfseries 80} no.~5, (2020) 409},
  \href{https://arxiv.org/abs/2001.05024}{{\ttfamily arXiv:2001.05024
  [hep-ph]}}.

\end{thebibliography}\endgroup

\end{document}